\documentclass[]{emulateapj}
\usepackage{epsfig, natbib, graphicx, color}

\input epsf 

\usepackage{color}

\newcommand{\Mpc}{\mbox{Mpc}}
\newcommand{\hMpc}{h^{-1}\mbox{Mpc}}
\newcommand{\msun}{M_\odot}

\newcommand{\bm}[1]{\mathbf{#1}}

\newcommand{\avg}[1]{\left\langle #1 \right\rangle}
\newcommand{\Var}{\mbox{Var}}

\newcommand{\bx}{\bm{x}}

\newcommand{\lin}{\bar \lambda_{in}}
\newcommand{\lout}{\lambda_{out}}

\newcommand{\qmin}{q_{min}}

\newcommand{\be}{\begin{equation}}
\newcommand{\ee}{\end{equation}}
\newcommand{\bea}{\begin{eqnarray}}
\newcommand{\eea}{\end{eqnarray}}

\newcommand{\bloc}{b_{local}}
\newcommand{\bmodel}{b_{model}}

\newcommand{\Rc}{R_{c}}

\shortauthors{Rozo et al.}
\shorttitle{Extrinsic Scatter of the Richness--Mass Relation}

\begin{document}
\title{Extrinsic Sources of Scatter in the Richness--Mass Relation of Galaxy Clusters}
\author{Eduardo Rozo\altaffilmark{1,2}, Eli Rykoff\altaffilmark{3}, Benjamin Koester\altaffilmark{4}, Brian Nord\altaffilmark{5,7}, 
Hao-Yi Wu\altaffilmark{6}, August Evrard\altaffilmark{5}, Risa Wechsler\altaffilmark{6}}
\altaffiltext{1}{Einstein Fellow, Department of Astronomy \& Astrophysics, The University of Chicago, Chicago, IL 60637.}
\altaffiltext{2}{Kavli Institute for Cosmological Physics, Chicago, IL 60637.}
\altaffiltext{3}{Lawrence Berkeley National Laboratory, Berkeley, CA 94720.}
\altaffiltext{4}{Department of Astronomy \& Astrophysics, The University of Chicago, Chicago, IL 60637.}
\altaffiltext{5}{Department of Physics, University of Michigan, Ann Arbor, MI 48109.}
\altaffiltext{6}{Kavli Institute for Particle Astrophysics and Cosmology, Physics Department, Stanford University, Stanford, CA 94305.}
\altaffiltext{7}{AGEP Fellow, Department of Physics, University of Michigan, Ann Arbor, MI 48109.}

\begin{abstract}
Maximizing the utility of upcoming photometric cluster surveys requires a thorough understanding of the richness--mass relation 
of galaxy clusters. We use Monte Carlo simulations to study the impact of various sources of observational scatter on this relation.  Cluster ellipticity, 
photometric errors, 
photometric redshift errors, and cluster-to-cluster variations in the properties of red-sequence galaxies contribute negligible noise.
Miscentering, however, can be important, and likely contributes to the scatter in the richness--mass relation of galaxy maxBCG
clusters at the low mass end, where centering is more difficult.
We also investigate 
the impact of projection effects under several empirically motivated assumptions about cluster environments. Using SDSS
data and the maxBCG cluster catalog, we demonstrate that variations in cluster environments can rarely
($\approx 1\%-5\%$ of the time) result in significant richness boosts.
Due to the steepness of the mass/richness function, 
the corresponding fraction of optically selected clusters that suffer from these projection effects is $\approx 5\%-15\%$.
We expect these numbers to be generic in magnitude, but a precise determination requires detailed, survey-specific modeling.

\end{abstract}

\section{Introduction}

The abundance of galaxy clusters as a function of mass has long been recognized as a powerful cosmological tool
\citep{peeblesetal89,evrard89}.
Clusters are highly complementary to other dark energy experiments \citep{cunhaetal09}, and, along with weak lensing tomography, 
cluster abundances can distinguish between dark energy and modiÞed gravity as the fundamental driver of our Universe's current phase
of accelerating expansion \citep{shapiroetal10}.
Unfortunately, cluster mass is not a direct observable, so
one is forced instead to compute the abundance of galaxy clusters as a function of observables that correlate with mass.
If the observable--mass relation is well understood, one can 
effectively recover the cosmological information inherent in the halo mass function.

Over the next several years, a variety of large scale photometric surveys such as the Dark Energy Survey 
(DES)\footnote{http://www.darkenergysurvey.org/}
and the Large Synoptic Survey Telescope (LSST)\footnote{http://www.lsst.org/lsst} 
are expected to observe large fractions of the sky to extremely faint magnitudes.  Using
a wide variety of techniques \citep[e.g.][]{gladdersyee00,milleretal05,koesteretal07,dongetal08,wenetal09,milkeraitisetal10,
haoetal10}, 
these surveys will identify hundreds of thousands of clusters
out to redshift, $z\sim 1$ and beyond, resulting in catalogs of
incredible statistical power.  To fully realize the promise of such catalogs, however, we must have a detailed
understanding of the richness--mass relation of these systems.
In particular, we must understand not only what the mean scaling between richness and mass is, 
but also how individual galaxy clusters scatter about this relation.   

In general, we expect there to be two distinct sources of noise which must be characterized,
namely {\it intrinsic} and {\it extrinsic} scatter.
By intrinsic scatter, we mean the fact that the
galaxy content of halos of a given mass will randomly vary from halo to halo, even if we could
unambiguously identify which galaxies belong to which halo with complete confidence.
In this work, we will simply assume that intrinsic scatter can be adequately modeled as Poisson.
This Poisson model is found to be a reasonable description of the scatter in the halo occupation
distribution of both dark matter halos \citep{kravtsovetal04} and simulated galaxies \citep[e.g.][]{berlindetal03,zhengetal05},
and is a key-component of halo model fits of galaxy clustering \citep[e.g][]{blakeetal08,zhengetal09,zehavietal10,tinkeretal10}.
We note, however, that recent work suggests the intrinsic scatter may in fact be significantly super-Poisson
in the large occupation limit \citep{boylan-kolchinetal10,wetzelwhite10,bushaetal10}.  If so, this can only
decrease the relative importance of the extrinsic sources of scatter considered here.

Let us turn then to the focus of this paper: extrinsic sources of scatter. In principle, there is a plethora of effects
that must be accounted for, such as halo triaxiality, photometric errors, projection effects, etc., so
the prospect of such a calibration is truly daunting.
In practice, however, not all of these possibilities are necessarily observationally
relevant.  {\it The goal of this paper is to identify all sources of extrinsic scatter that are observationally
relevant.}   Having identified the relevant sources of scatter, we intend to return in a future work to the
problem of modeling the relevant sources of scatter quantitatively.   Note that if a source of extrinsic scatter is negligible
relative to Poisson intrinsic scatter,  it is also negligible relative to super-Poisson scatter, so our analysis
is conservative in that sense.

To identify which sources of measurement noise are observationally relevant we rely on a Monte Carlo
approach: using analytic models, we simulate galaxy clusters under 
a variety of different assumptions about extrinsic sources of noise, and then
estimate their richness to determine whether the richness--mass relation was affected at a
significant level (where ``significant'' is quantitatively defined below).
While this method lacks the
sophistication of N-body simulations, it has the significant advantage of permitting control of every detail
of the simulated clusters.  More specifically, we can turn on the sources of extrinsic scatter  one at a time
in order to identify those that are observationally relevant.
Moreover, it gives us the ability to very quickly simulate hundreds of thousands of clusters, which is necessary for
us to accomplish our goals.
We believe this Monte Carlo approach should suffice for our stated goal of identifying relevant sources of scatter,
though clearly further work will be required to quantitatively characterize those sources of scatter that are 
observationally relevant.

This paper is the second in a series of papers that have as their final goal the development of a fully optimized
richness estimator that is both qualitatively and quantitatively understood.  The first of these papers  
\citep[][henceforth referred to as paper I]{rozoetal09b} introduced
the general matched filter formalism we use to estimate cluster richness,
and tuned it to minimize the scatter in X-ray luminosity of galaxy clusters at fixed richness
\citep[similar approaches towards richness estimation and cluster finding can be found in][]{kepneretal99,
whitekochanek02,kochaneketal03,dongetal08}.
Here, we identify those sources of extrinsic scatter that can impact the observed richness--mass relation,
while a companion paper (paper III) considers various modifications to the richness estimator $\lambda$ in an effort to further
improve the fidelity with which cluster richness traces mass \citep[][henceforth referred to as paper III]{rykoffetal10}.
Our final richness estimator is extremely robust, and we believe is close to being fully optimal
for counting red-sequence galaxies from photometric data.
We intend to follow these two papers with an additional quantitative study of the sources of extrinsic
scatter identified in this work.

The layout of this paper is as follows: in section \ref{sec:lambda} we quickly review the richness estimator $\lambda$ from paper I, while
section \ref{sec:method} details our method for generating Monte Carlo realizations of galaxy clusters.  Section \ref{sec:results}
explores a wide variety of extrinsic sources of scatter, and section \ref{sec:proj} considers the impact
of projections effects on cluster richness for a variety of empirically motivated assumptions about the environment of galaxy clusters.
Section \ref{sec:summary} summarizes our results.  Appendix \ref{app:scatter} details the reasoning behind
the definition adopted in the main body of the work as to what constitutes an observationally relevant source of scatter.
Unless otherwise stated, all of our calculations assume a fiducial flat $\Lambda$CDM cosmology with $h=0.7$ and $\Omega_m=0.25$.


\section{The Richness Measure $\lambda$}
\label{sec:lambda}

We begin by summarizing the algorithm behind the matched filter richness
$\lambda$ originally proposed in paper I.   Let $i$ index all galaxies around a
putative cluster center, and $y_i$ be a random variable
such that $y_i=1$ if a galaxy is a member of a galaxy clusters, and
$y_i=0$ otherwise.  The richness is defined as the total number of
cluster galaxies
\begin{equation}
\lambda=\sum_i y_i.
\end{equation}
Define $p_i$ as the membership probability of galaxy $i$, so that
$P(y_i=1)=p_i$ and $P(y_i=0)=1-p_i$.  The mean and variance of the richness $\lambda$
are given by
\begin{eqnarray}
\avg{\lambda} & = & \sum_i \avg{y_i} = \sum_i p_i \label{eq:mean} \\
\Var(\lambda) & = & \sum_i \Var(y_i) = \sum_i p_i(1-p_i) \label{eq:lambdavar}.
\end{eqnarray}
If the membership probabilities $p_i$ are known, then equation \ref{eq:mean} 
can be used to define the richness estimator $\hat \lambda$ via $\hat \lambda=\avg{\lambda}$,
while equation \ref{eq:lambdavar} would correspond to the statistical uncertainty in $\hat \lambda$.

Expanding the product in the expression for the variance, we find
\begin{equation}
\Var(\lambda) = \avg{\lambda}(1-\bar p)
\end{equation}
where $\bar p$ is the mean membership probability,
\begin{equation}
\bar p = \frac{\sum_i p_i^2}{\sum_i p_i} = \frac{1}{\avg{\lambda}}\sum_i p_i^2.
\end{equation}
This implies that as long as the mean membership probability is close to unity,
the statistical uncertainty in the richness is significantly smaller than Poisson.

The membership probabilities are estimated as follows:
let $\bm{x}$ be a vector describing the observable properties of a galaxy
(e.g. galaxy color, magnitude, and position).  We model the projected galaxy
distribution around clusters as a sum $S(\bx)=\lambda
u(\bx|\lambda)+b(\bx)$, 
where $\lambda$ is the number of cluster galaxies, $u(\bx|\lambda)$ is the
number density profile of cluster galaxies normalized to unity, and $b(\bx)$ is the density
of background (i.e. non-member) galaxies.  The probability that a galaxy found near a cluster
is actually a cluster member is simply
\begin{equation}
p(\bx) = \frac{\lambda u(\bx|\lambda)}{\lambda u(\bx|\lambda)+b(\bx)}.
\end{equation}

Inserting these probabilities back into equation \ref{eq:mean} we arrive at
\begin{equation}
\lambda  = \sum p(\bx|\lambda) = \sum_{R<\Rc(\lambda)} \frac{\lambda u(\bx|\lambda)}{\lambda u(\bx|\lambda)+b(\bx)}.
\label{eqn:lambdadef}
\end{equation}
Equation \ref{eqn:lambdadef} can be solved for the value of $\lambda$, which in turn defines our richness
estimator.
In principle, the sum should extend over all galaxies.  In practice, one 
needs to add over all galaxies within some cutoff radius $R_c$ and above some
luminosity cut $L_{cut}$, for which we set $\Rc=1\ \Mpc$ and $L_{cut}=0.2L_*$
unless otherwise noted.  The choice of a fixed metric aperture is purely for simplicity.
Paper III has a detailed discussion of the impact of the radial aperture on the richness--mass
relation, and optimizes the radial aperture so as to minimize the scatter in $L_X$ at fixed richness.

We consider three observable galaxy properties: $R$, the
projected distance from the cluster center; $m$, the galaxy $i$-band magnitude;
and $c$, the galaxy $g-r$ color, as appropriate for low redshift galaxy clusters.
We adopt a separable filter function
\begin{equation}
u(\bx) = [2\pi R \Sigma(R)]\phi(m)G(c)
\end{equation}
where $\Sigma(R)$ is the two dimensional cluster galaxy density profile,
$\phi(m)$ is the cluster luminosity function (expressed in apparent
magnitudes), and $G(c)$ is color distribution of cluster galaxies. The
pre-factor $2\pi R$ in front of $\Sigma(R)$ accounts for the fact that given
$\Sigma(R)$, the radial probability density distribution is $2\pi R
\Sigma(R)$.  
We summarize each of these filters below.


\subsection{The Radial Filter}

We assume cluster galaxies follow an NFW profile~\citep{navarro_etal95}.
The corresponding two-dimensional surface density profile is \citep{bartelmann96}
\begin{equation}
\Sigma(R) \propto \frac{1}{(R/R_s)^2-1}f(R/R_s)
\label{eqn:radfilter}
\end{equation}
where $R_s$ is the characteristic scale radius, and
\begin{equation}
f(x) = 1-\frac{2}{\sqrt{x^2-1}}\tan^{-1}\sqrt{ \frac{x-1}{x+1} }.
\end{equation}
This formula assumes $x>1$. For $x<1$, one uses the identity $\tan^{-1}(ix) = i \tanh(x)$.  
As in paper I, we set $R_s=0.15\,\hMpc$.  While the profile is singular
as $R\rightarrow 0$, the membership probability remains finite with $p\rightarrow 1$
as $R\rightarrow 0$.  The filter 
$\Sigma(R)$ is normalized to unity within our chosen fixed metric aperture $\Rc=1\ \Mpc$,
\begin{equation}
\label{eqn:radnorm}
1 = \int_0^{\Rc} dR\ 2\pi R\Sigma(R).
\end{equation}
%


\subsection{The Luminosity Filter}

The luminosity distribution of maxBCG clusters is well represented by a
Schechter function~\citep[e.g.][]{hansenetal07}, which we write as
\begin{equation}
\phi(m) = C 10^{-0.4(m-m_*)(\alpha+1)}\exp\left(-10^{-0.4(m-m_*)}\right)
\label{eqn:lumfilter}
\end{equation}
where $C=0.4\ln(10)\phi_*$.  The normalization $\phi_*$ is fixed by requiring that $\phi$
be normalized to unity above a luminosity cut $L_{cut}$, which we set to $L_{cut}=0.2L_*$.
This is fainter than the cut adopted  in paper I, but matches the final cut we
adopt in paper III.  
We also set $\alpha=0.8$, and we calculate $m_*$ using passively evolved stellar population
models \citep[see][for details]{koesteretal07}.  An accurate polynomial interpolation for $i_{cut}(z)$ 
is
\be
i_{cut}= 19.605+2.327x + 0.205x^2 + 0.202x^3
\ee
where $x=\ln(1+\delta_z)$ and $\delta_z=(z-0.2)/0.2$.  This fitting function is accurate at the level of $\Delta i_{cut}=0.002$
over the redshift range $0.1\leq z \leq 0.3$.  

\subsection{The Color Filter}

For a color filter, we assume red-sequence galaxies have a Gaussian color
distribution with intrinsic dispersion $\sigma_{\mathrm{int}} = 0.05\,\mathrm{mag}$.  The
corresponding color filter, $G(c)$ is
\begin{equation}
\label{eqn:color}
G(c|z) =\frac{1}{\sqrt{2\pi}\sigma}\exp \left[ \frac{(c-\avg{c|z})^2}{2\sigma^2} \right],
\end{equation}
where $c=g-r$ is the color of interest, $\avg{c|z}$ is the mean of the Gaussian
color distribution of early type galaxies at redshift $z$, and $\sigma$ is the
width of the distribution.  The mean color $\avg{c|z}=0.625+3.149z$ was
determined by matching maxBCG cluster members to the SDSS LRG
\citep{eisensteinetal01} and MAIN \citep{straussetal02} spectroscopic galaxy
samples.  The net dispersion $\sigma$ is taken to be the sum in quadrature of
the intrinsic color dispersion $\sigma_{int}=0.05$ and the estimated color
error $\sigma_c$ of each individual galaxy.


\subsection{Background Model}
\label{sec:background}

The last necessary ingredient for estimating $\lambda$ is a background model.
We assume the background galaxy density is constant in space, so that
$b(\bx)=2\pi R \bar \Sigma_g(m_i,c)$ where $\bar\Sigma_g(m_i,c)$ is the galaxy
density as a function of galaxy $i-$band magnitude and $g-r$ color. 
In paper III 
we estimate $\bar\Sigma_g$ by binning SDSS galaxies in color--magnitude space
using a cloud-in-cell (CIC) algorithm~\citep[e.g.][]{hockney81}.  We use
those results to define our background model at every point through linear
interpolation.    Further details of how the background density model is constructed
can be found in paper III.  For our purposes, the key is that we have
an empirically determined function $\bar\Sigma_g(m_i,c)$ that returns the
mean galaxy density of the universe as a function of $i$-band magnitudes
and $g-r$ color.


\section{Method}
\label{sec:method}

Our idea is simple: using the filters $\Sigma(R)$, $\Phi(m)$, and $G(g-r)$ ,
we generate Monte Carlo realizations of a single cluster, and measure the
scatter in richness.  We can then repeat the experiment including an extrinsic
source of scatter, and determine whether the newly introduced effects is observationally
relevant or not.

We simulate the galaxy
density field around a galaxy cluster out to a $3\ \Mpc$ radius,
comfortably larger than the $1\ \Mpc$ aperture used to estimate cluster 
richness.  To generate the galaxy fields, 
we first select the expectation value $\lin$ for the number of
red sequence galaxies in a cluster within a $1\ \Mpc$ radius.
Using the radial filter $\Sigma(R)$, we extrapolate in radius to 
compute the expected number of red sequence galaxies within $3\ \Mpc$,
which we label $\bar N_3$.
The number of galaxies assigned to the cluster is a Poisson
realization of mean $\bar N_3$.
Each cluster galaxy is assigned a radius, angle, 
magnitude, and color, by randomly sampling the filters
$\Sigma(R)$, $\Phi(m)$, and $G(c)$.   The brightest galaxy is always
placed at the center of the cluster.  
If $\lin$ is small, it is possible for no galaxies to be bright enough to pass the magnitude cut.
Since observationally we are restricted to systems with at least one bright galaxy, whenever
this happens we simply add a central galaxy.  Note that means that when $\lin \rightarrow 1$,
we expect large biases in the sampled of detected clusters simply because one misses
all systems with no galaxies.

Once cluster galaxies are in place, we use a similar
procedure to populate our sky patch with non-cluster galaxies. Given a background
model $b(R,m,c)$, we compute the expected
number of such galaxies, draw the number of background galaxies from a Poisson distribution
of the appropriate mean, and then assign to every galaxy a radius, angle, magnitude, 
and color using the background filters.  Note that the background model used to generate our Monte
Carlo simulations need not be the same as the background model employed to estimate
$\lambda$.  
Throughout section \ref{sec:results}, however, we will use the density model
from section \ref{sec:background} both for populating our simulations with non
cluster galaxies, and for estimating richness.
To convert the observed mean galaxy density from $\mbox{gal}/\mbox{deg}^2$
to $\mbox{gal}/\Mpc^2$, we assume a flat $\Lambda$CDM
cosmology with $\Omega_m=0.3$ and $h=0.7$.

Once a patch has been generated as described above, we estimate its richness, which we henceforth
refer to as $\lout$.  The richness is estimated as described in section \ref{sec:lambda}.
To measure the statistical properties of $\lout$, we repeat this procedure a minimum of $400$
times.  When confronted with noisy realizations (e.g. when considering miscentering), we increase the number of
samples up to 10,000 realizations (per choice of miscentering parameters).
Finally, we emphasize that 
since $\lin$ is a deterministic function of the mass (e.g. $\lin \propto M^\alpha$), holding
$\lin$ constant is equivalent to holding halo mass constant.  That is, the scatter we measure
is precisely the scatter in richness at fixed mass, $\sigma_{\ln \lambda|M}$.
Throughout, we use
the word ``scatter'' to signify the standard deviation of $\ln \lout$ at fixed $\lin$.

It is obvious from the above description that in these Monte Carlo realizations there is no allowance
for contamination of the cluster field by correlated galaxies.  We address this
difficulty in section \ref{sec:proj}, where we change the background model used to populate
cluster fields with non-cluster galaxies.
For our purposes, the most relevant result concerning projection effects is that they
are rare, but severe.  That is, projection effects don't really broaden the peak of the
probability distribution $P(\lout|\lin)$.  Rather, they build a small non-gaussian where
$\lout$ is much larger than expected. 
Throughout section \ref{sec:results}, we focus exclusively
on the central component of the distribution $P(\lout|\lin)$.

Using the method described above to generate Monte Carlo realization of galaxy clusters, 
we can measure
the distribution $P(\lout|\lin)$.  In most cases, we focus exclusively on the mean and scatter
of this distribution.  The two obvious questions that can be addressed with such data are:
1) is our richness estimator biased? 2) is the scatter of our richness estimator consistent with Poisson?

Concerning the first question, we emphasize that 
{\it biases in our richness estimator are irrelevant}.  The richness measure $\lambda$ is only meant to
be interpreted as an observational quantity that scales with mass with little scatter.  The scaling or richness
with both mass and redshift 
needs to be empirically calibrated regardless of whether $\lout$ is biased
relative to $\lin$ or not.  Thus, biases in $\avg{\lout|\lin}$ are of no practical consequence.  It is the
scatter in $\lout$ that we are primarily concerned about.

Finally, we need to determine how much extrinsic scatter can we tolerate before the extrinsic scatter
becomes observationally relevant, i.e. how far can we deviate from Poisson scatter.
Here, we focus on whether the scatter in $\lout$ can be modeled
as Poisson for the purposes of the Dark Energy Survey (DES).  In appendix \ref{app:scatter}, we 
demonstrate that differences between the true and predicted scatter of the richness--mass relation are irrelevant
so long as these differences are about $5\%$ or less,  (i.e. $\Delta\sigma_{\ln \lambda} \lesssim 0.05$).
Thus, for this work, we will say that a source of 
scatter is observationally irrelevant whenever $\Delta\sigma_{\ln \lambda} \leq 0.05$.\footnote{
To estimate $\sigma_{\ln \lambda}$ for a Poisson distribution, we use $10^5$ Poisson realizations
and numerically estimate $\sigma_{\ln \lambda}$.  As with the real data, if a realization results in
no galaxies, we set $\lout=1$ instead.}


\section{Sources of Scatter for the Richness--Mass Relation of Galaxy Clusters}
\label{sec:results}

We wish to determine what sources of statistical and/or systematic uncertainty can significantly impact
the observed richness--mass relation.  The effects we consider are halo triaxiality,
cluster-to-cluster scatter in the properties of ridgeline galaxies, photometric errors, photometric
redshift errors, and cluster miscentering.
Before we proceed, however, we must set a baseline, and determine what the
richness--mass relation of galaxy clusters is in the absence of any such additional
sources of noise.

\subsection{The Intrinsic Scatter of the Richness--Mass Relation}
\label{sec:intscat}

\begin{figure}[t]
\epsscale{1.2}
\plotone{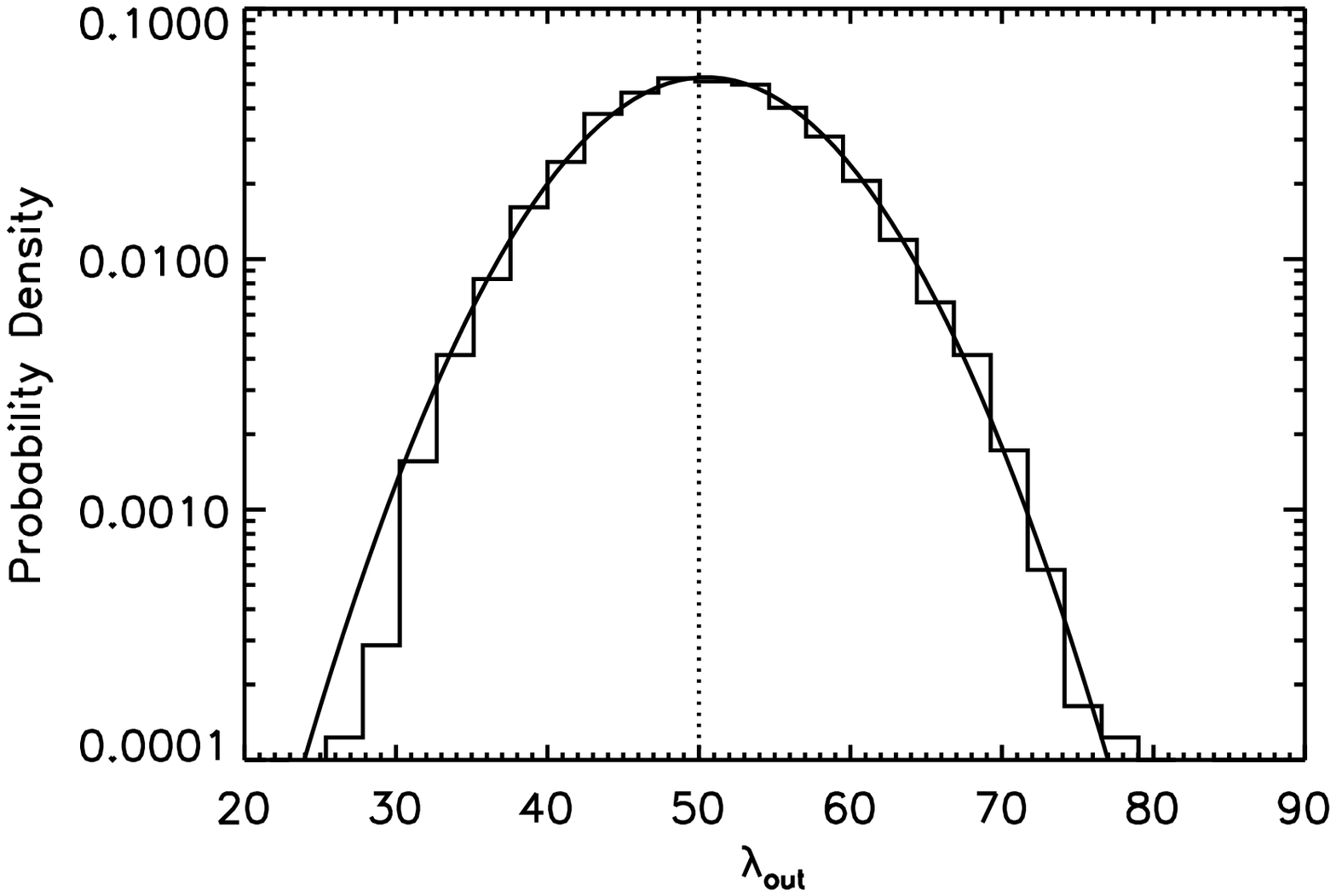}
\plotone{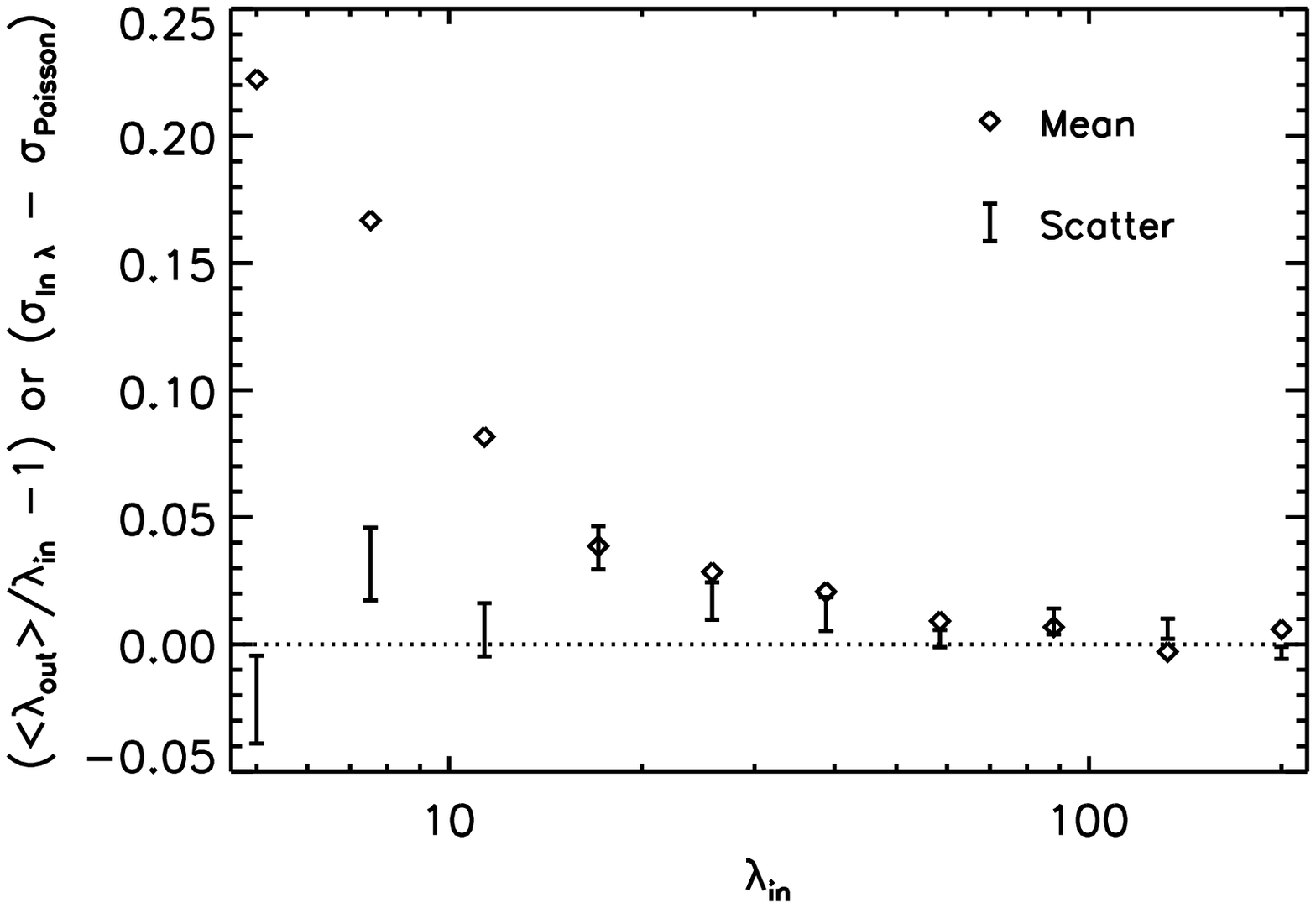}
\caption{{\it Top panel:} The distribution of the estimated richness $\lout$ for $10^4$ independent realizations
of a single cluster with $\lin=50$.   The distribution has a slight ($1\%$) bias in the mean, and the
scatter is very well approximated by Poisson.  Note that since the richness--mass relation needs to be empirically calibrated,
a bias in the mean is not of importance to us.
 {\it Bottom panel:} 
The first two moments of the $\lout-\lin$ relation as a function of richness.
Diamonds show the bias of our richness estimator $\avg{\lout}/\lin-1$, while the points with error bars show the deviation
$\Delta\sigma_{\ln \lout}$ of the observed scatter in $\ln \lout$ from the Poisson expectation.
For $\lin \geq 20$, our richness estimates are slightly biased,  and the scatter can be approximated as Poisson.}
\label{fig:lambda_dist}
\end{figure}

The top panel in Figure \ref{fig:lambda_dist} 
shows the distribution of $\lout$ obtained from $10^4$ realizations of a cluster with $\lin=50$ galaxies. 
The solid curve is the best-fit Gaussian, and the vertical dotted line is the input richness.
Two things are evident from this figure: first, our richness estimator is nearly unbiased,
and second, the distribution $P(\lout|\lin)$ can be adequately modeled as Poisson.
Of course, readers will be quick to note that the
actual number of cluster galaxies we place in our $3\ \Mpc$ field is itself drawn from a Poisson distribution.
Are we engaging in circular reasoning?  


The answer is only partially.
Consider the process used to create and test the Monte Carlo simulations: given $\lin$, 
we draw a Poisson realization for the number of galaxies in the cluster field,
in accordance with naive expectation for the {\it intrinsic} scatter of the richness--mass relation.
If the reasoning were circular, the exercise would conclude with this stage.  
In our simulations, however, we use Poisson statistics to populate a $3\ \Mpc$
cluster field with galaxies,  we then randomly add background galaxies, and finally, we estimate the richness
within a $1\ \Mpc$ aperture.  This whole procedure must necessarily introduce some amount of
measurement error (for instance, $\lambda$ is not an integer), 
but it is readily apparent from our results that this extra noise is negligible relative to the expected
intrinsic scatter.  Thus, the scatter we recover is Poisson both because the intrinsic scatter is Poisson,
and because the measurement error associated with estimating richness is negligible, as we had anticipated
from equation \ref{eq:lambdavar}.

The bottom panel in Figure \ref{fig:lambda_dist} explores the extent to which our conclusions depend on the richness of the cluster under consideration.  
The figure shows the bias in the mean (diamonds) as well as the deviation from the Poisson expectation for the standard deviation
(points with error bars)
as a function of the input richness $\lin$.  As we can see, our above
conclusions are valid at richnesses $\lambda \gtrsim 15$, though
measurement error does become more important with decreasing cluster richness.  The bias at low richness is due to the fact that we demand
a central galaxy to always be present.


\subsection{Halo Triaxiality}
\label{sec:ellipticity}

Halos are known to be triaxial,
with halo triaxiality depending on both halo mass and redshift \citep[e.g.][]{bettetal07,kasunevrard05,knebewiessner06,
shawetal06,allgoodetal06,jingsuto02}.
This has two important consequences:
first, the projected galaxy density of a halo is not circularly symmetric, and second, the amplitude of the density
field is itself modulated by the line of sight projection.  We explore both effects, beginning
with the impact of non-circular symmetry.   
We modify our simulations as follows: first, we select the minimum projected axis ratio $\qmin$ that an elliptical cluster can
have in our realizations.   The projected axis ratio $q$ of each realization is drawn uniformly from the range
$[\qmin,1]$.  While not realistic, this certainly
suffices for the purposes of determining whether scatter in the projected ellipticity of a halo can introduce
significant scatter in the richness--mass relation.  For each mock realization, we draw a different axis
ratio $q$, and randomly place galaxies according to the corresponding elliptical halo profile $u(\rho)$
where $\rho$ is now an elliptical coordinate, $\rho^2=x^2+q^2y^2$.    All other aspects of our simulation
remain unchanged.

We find that cluster ellipticity impacts the richness--mass relation only at the $\lesssim 2\%$ level in the mean,
and that the scatter can be described as Poisson at comparable accuracy for $q_{min}\geq 0.5$.  
Consequently, variance in the ellipticity of the galaxy distribution does not significantly
impact the richness--mass relation of galaxy clusters.  This result 
emphasizes an important distinction that often gets overlooked: using radial
filters is {\it not} equivalent to assuming spherical symmetry.  For instance, lining up galaxies while preserving their radial
distribution has no impact on a cluster's richness as defined in section \ref{sec:lambda}.  
Thus, it is not surprising that the projected ellipticity of the galaxy distribution of halo has little effect on the richness--mass relation.

We now turn to the line of sight modulation of the amplitude of the galaxy density field.  We adopt a simple
model of the form
\be
\Sigma_{g,cluster}(R) = N \lin u(R)
\label{eq:tri}
\ee
where $u(R)$ is the cluster profile normalized to unity, and 
$N$ is a normalization constant that depends on the axis ratios $q_1$ and $q_2$ of the triaxial halo and 
on the line of sight $\bm{\hat n}$ along which the halo is projected.  Given the statistical properties
of the triaxial halo population, one can compute the corresponding
probability density $P(N)$.  However,
because our goal is only to determine whether halo triaxiality can significantly impact the scatter of the richness--mass
relation, we consider instead a simple 
Gaussian 
distribution for $N$, and explore how the richness--mass relation depends on the standard deviation of $N$, 
which we denote $\Delta N$.  To estimate a typical value for $\Delta N$, we rely on the results 
of \citet{rozoetal07c}, who computed
$N(q_1,q_2,\bm{\hat n})$ for the case of a tri-axial singular isothermal ellipsoid.  Assuming the axis
$q_1$ and $q_2$ are drawn from a uniform distribution $q_1\in[0.5,1]$ and $q_2\in[q_1,1.0]$, and that the halo
orientation $\bm{\hat n}$ is random, we find $\Delta N=0.085$.


\begin{figure}[t]
\epsscale{1.2}
\plotone{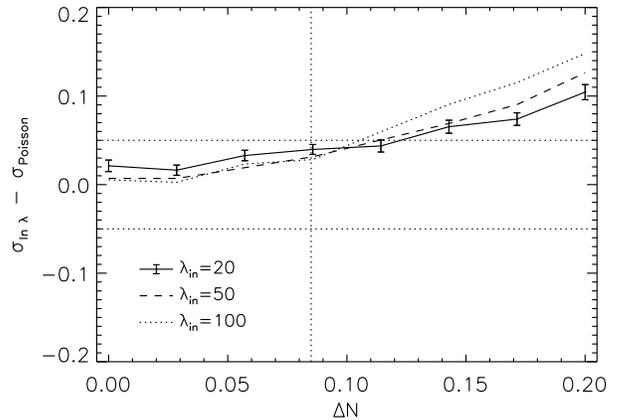}
\caption{Bias
and deviation from the Poisson expectation of the scatter of the richness--mass relation as a function of $\Delta N$, the standard
deviation of the surface density amplitude for triaxial halos.  The vertical dotted line marks a rough estimate for 
the expected value for $\Delta N$ for triaxial halos.
We find halo-triaxiality can increase the scatter of the richness--mass relation by $\Delta\sigma_{\ln \lambda}=0.04$, 
which is border-line important.  For clarity, we only show error bars for the $\lin=20$ case, but the errors are comparable
for the different $\lin$ values.
}
\label{fig:tri}
\end{figure} 


Figure \ref{fig:tri} shows the difference
between the observed scatter and the Poisson expectation as a function of $\Delta N$ 
for three different richness clusters as labelled.  
The bias in the mean is very nearly zero, and is therefore not shown.  Evidently, large $\Delta N$ values can contribute significantly
to the observed scatter.
For the expected level of halo-triaxiality (vertical dotted line), the impact of this effect on the richness--mass
relation is border-line significant.


\subsection{Cluster-to-Cluster Variance of Ridgeline Properties}
\label{sec:ridgelinescat}

In a recent work, \citet{haoetal09} found cluster-to-cluster variations of order 
$0.05$ magnitudes in the mean color of ridgeline cluster galaxies at fixed redshift.
Unfortunately,
they did not determine whether such fluctuations were intrinsic, or simply reflected
measurement error.  If the former dominates, 
one needs to consider whether such intrinsic fluctuations could impact the richness--mass
relation of galaxy clusters.

In our original simulations, the color of cluster galaxies is randomly
drawn from a Gaussian distribution of mean $\bar c$ and standard deviation $\sigma_c=0.05$.
We include variations in the properties of ridgeline galaxies by setting
the mean color of cluster galaxies to $\bar c = \bar c_0+\Delta c$ where
$\bar c_0$ is the mean color of cluster galaxies over all clusters, and $\Delta c$ is a random
Gaussian color offset of zero mean and standard deviation $\sigma_{\bar c}$.

We find that the scatter in ridgeline properties has no impact on the mean or scatter of the
richness--mass relation at the $1\%$ level. 
This robustness reflects the fact that $\lambda$
is not particularly sensitive to the precise location of the central peak in the color filter (see papers I and III for an 
extended discussion),
so that small differences between the color distribution used to generate cluster galaxies and the actual
filter employed in estimating cluster richness have only a modest impact on the final cluster richness estimate.
This is also good news for the purposes of extending this work to higher redshifts, since any intrinsic 
cluster-to-cluster variation in the properties of ridgeline galaxies would likely increase with
increasing redshift.  In paper III, we explicitly consider the sensitivity of $\lambda$
to the details of the red-sequence model.  More specifically, we demonstrate that $\lambda$
is robust to whether we employ our red-sequence model, or whether we empirically fit for
the red-sequence parameters on a cluster-by-cluster basis.  This is consistent with
the analysis presented here that cluster-to-cluster variance in ridgeline properties
is not an important source of scatter in the richness--mass relation.


\subsection{Photometric Errors}
\label{sec:photo}

Photometric errors impact a cluster's richness estimate in two ways: they can
scatter galaxies across the luminosity cut used to count galaxies, and
they can scatter the location of a galaxy in color space.
We test the impact of each of these effects in turn.

To include photometric errors in our simulations, we compute the median error
$\sigma_i(i)$ and $\sigma_{g-r}(i)$ on $i$ and $g-r$ as a function of $i$ in
the SDSS.  For each galaxy in our simulation, we then set $i_{obs}=i_{true}+\Delta i$
where $\Delta i$ is a random Gaussian offset of zero mean and standard deviation 
$\sigma_i(i_{true})$.  Finally, the galaxy is assigned a photometric error $\sigma_i(i_{obs})$.
A similar operation can be performed with $g-r$.  
In order to disentangle the impact
of photometric errors in $i$ with those in $g-r$, we consider simulations in which $i$ band
magnitudes are subject to photometric errors, but $g-r$ is not, as well as the converse case.

We parameterize the photometric error function $\sigma_i(i)$ and $\sigma_{g-r}(i)$ as
\bea
\sigma_i(i)  & = & \sigma_{i,20}\exp \left( 0.56(i-20) \right) \\
\sigma_{g-r}(i) & = & \sigma_{g-r,20}\exp \left( 0.70(i-20) \right).
\eea
The pivot point $i=20$ is chosen simply as a convenient reference magnitude for galaxies at
redshift $z\approx 0.25$ brighter than $0.2L_*$.
These parameterization provide reasonable fits to the SDSS data.  In our simulations,
we vary the amplitude parameters in the range $\sigma_{i,20}\in[0,0.2]$ and $\sigma_{g-r,20}\in[0,0.4]$.
The amplitudes in the SDSS are $\sigma_{i,20}=0.04$ and $\sigma_{g-r,20}=0.11$.  We find that over
the ranges probed --- which extend significantly beyond the precision of SDSS photometry --- 
photometric errors do not affect the scatter in richness--mass relation of galaxy clusters
in a significant way.  Not surprisingly, sufficiently large photometric errors can impact the mean cluster
richness, but doing so requires a photometric uncertainty $\sigma_{i,20} \gtrsim 0.1$, significantly
larger than the $0.04$ error from SDSS.  Thus, the photometric calibration in SDSS has a negligible
impact on the richness--mass relation of galaxy clusters.  This conclusion will only be strengthen in
future surveys, where photometric uncertainties will be further reduced.
This is also consistent with the tests we perform in paper III, where we demonstrate that $\lambda$
is only modestly affected by errors of up to $0.05$ magnitudes in the magnitude cut corresponding
to our luminosity threshold.


\subsection{Photometric Redshift Errors}
\label{sec:photozerr}

Our richness estimate $\lambda$ is the number of red-sequence galaxies brighter than a given luminosity cut and within a
specified radial cut.  However, the physical radial offset and luminosity of a galaxy depends on the
cluster's redshift, as does the color of the red-sequence.  Consequently,
scatter in a cluster's redshift can produce scatter in the richness--mass
relation.  To test whether this is significant,
we perform Monte Carlo realizations as described in section
\ref{sec:method}.  We then assume that the cluster is assigned an observed redshift
$z_{obs}=z_{true}+\Delta z$, where $\Delta z$ is a random Gaussian offset of zero mean
and standard deviation $\sigma_z$.   The assigned luminosity of every galaxy is then rescaled
by the square of the luminosity distance ratio between the true and observed redshift.  The radial
locations of the galaxies
are also scaled by the angular diameter distance ratio. The cluster richness is then estimated as usual.


\begin{figure}[t]
\epsscale{1.2}
\plotone{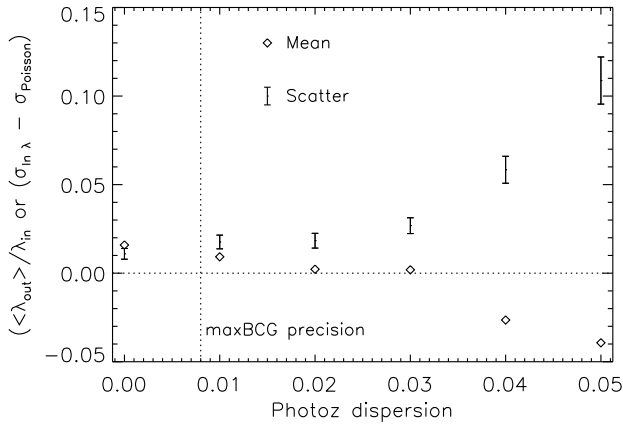}
\caption{Mean and scatter of the richness--mass relation as a function of photometric redshift
errors.  Photometric redshift errors need to be controlled at the $\Delta z \lesssim 0.02$ level
for them to have a negligible impact on the richness--mass relation.  This value is comfortably
above the estimated photometric redshift error of maxBCG clusters (vertical dotted line).}
\label{fig:photoz}
\end{figure} 


Figure \ref{fig:photoz} shows how the mean and scatter of the richness--mass relation depends
on the standard deviation of the photometric redshift errors $\sigma_z$.  What is remarkable
about this figure is how fast redshift errors go from irrelevant to significant. Below $\sigma_z \approx 0.02$,
photometric redshift errors are not very important, but their contribution to the total scatter 
rapidly increases with $\sigma_z$
beyond this point.  Thus, accurate photometric redshift estimates are necessary in order
to avoid significantly increasing the scatter of the richness--mass relation. 
For reference, the photometric redshift accuracy of SDSS maxBCG clusters is 
$\sigma_z\approx 0.008$, so cluster photometric redshift
errors are not a significant systematic in this case. 

We can easily understand why photometric redshift errors can become so important so quickly.
If one takes a galaxy at the median redshift of the cluster
sample, and displaces it in redshift by $\Delta z$, the corresponding magnitude change is roughly
$\Delta m = 10\Delta z$, so a redshift error of $0.02$ changes the apparent luminosity
of a cluster galaxy by $0.2$ magnitudes, a relatively large amount.  Moreover, unlike
photometric errors, all galaxies step in unison, and the effect of this scatter is not down-weighted
by root-N statistics.

The results of this section can also speak to the importance of uniform photometric
calibration when characterizing the scatter in the richness--mass relation.  As discussed
above, a photometric redshift error of $0.02$ corresponds to a photometric shift of $0.2$ magnitudes;
at this point, photometric redshift errors become important.  Consequently, in order for
the scatter of the richness--mass relation to remain unaffected by variations in the photometric
calibration of a survey, the latter needs to be controlled at the level of $0.2$ magnitudes.


\subsection{Cluster Miscentering}
\label{sec:miscentering}

We consider the impact that cluster miscentering --- i.e. the possibility that the cluster center chosen by
an observer may be offset from the true cluster center --- can have 
on the richness--mass relation.
We adopt the mis-centering parameterization used in \citet{johnstonetal07}
to characterize the centering properties of the maxBCG cluster-finding algorithm and explore how
our richness measure depends on these quantities.   The two parameters of interest are the probability
$p$ that a cluster be correctly centered, and the standard deviation $\sigma_R$ characterizing the
distribution of random offsets for mis-centered clusters (see below).

We incorporate mis-centering in our simulations as follows: for each cluster realization, we
randomly determine whether the cluster is mis-centered or not based on its centering probability. 
If the cluster is mis-centered, we  draw a random offset vector by randomly selecting a
position angle, and then randomly sampling the offset along the corresponding
axis from a Gaussian of zero mean and standard deviation $\sigma_R$.
We then 
measure the richness $\lambda$ about this new cluster center. 
Note that we are not enforcing the cluster
center to fall on a cluster galaxy.  This procedure 
allows us to vary the mis-centering parameters
in a smooth fashion in order to explore the sensitivity of our results to the input parameters.


\begin{figure}[t]
\epsscale{1.2}
\plotone{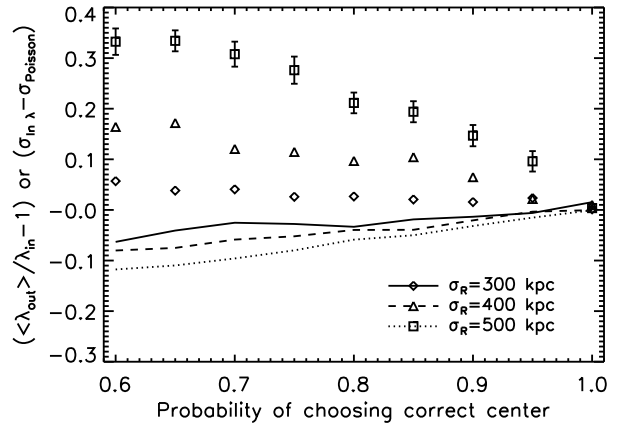}
\caption{Mean and scatter of the richness--mass relation as a function of mis-centering parameters.
Isolated symbols always track the scatter data, while the curves trace the mean data.
The parameter $\sigma_R$ is the standard deviation of the
random displacement vector applied when clusters are mis-centered.  Not surprisingly, mis-centering can severely
impact both the mean and the scatter of the richness--mass relation if mis-centering offsets are comparable to
the aperture used for estimating richness ($1\ \Mpc$ in our simulations).}
\label{fig:miscentering}
\end{figure} 


Figure \ref{fig:miscentering} shows how the mean (curves) and scatter (symbols) 
of the richness--mass relation depends
on the mis-centering parameters for a variety of $\sigma_R$ values: $\sigma_R=0.1\ \Mpc$ (solid, diamonds),
$\sigma_R=0.3\ \Mpc$ (dashed, triangles), and $\sigma_R=0.5\ \Mpc$ (dotted, squares).
The horizontal axis is the probability $p$ that a cluster be correctly centered.
Not surprisingly, when the mis-centering offset parameter
$\sigma_R$ is comparable to the $\lambda$-aperture
($\sigma_R\gtrsim R_\lambda/2$), richnesses are systematically underestimated and the scatter of the
richness--mass relation is dramatically increased.    Importantly, however, notice that miscentering
``turns-o'' remarkably fast.  For $p\approx 0.85$ and $\sigma_R/R \lesssim 0.4$, miscentering does not
appear to be an important systematic, but setting $p\approx 0.75$ and $\sigma_R/R=0.5$ significantly
increases the scatter.  This is an important feature that we will return to in Appendix B of paper III.

What does this imply for the scatter in richness at fixed mass for maxBCG clusters.  At $N_{200}\approx 25\ (50)$,
the miscentering parameter $p\approx 0.7 (0.8)$
and $\sigma_R\approx 0.4\ \hMpc \approx 0.57\ \Mpc$ 
\citep{johnstonetal07,hilbertwhite10}.  The richness $N_{200}=25 (50)$ corresponds roughly to $\lambda= 30 (60)$.
Using our optimal richness estimator, the corresponding apertures are $R_c = 1.1\ (1.3)\ \Mpc$, and therefore
the ratio $\sigma_R/R_c \approx 0.5\ (0.4)$.  Assuming intrinsic Poisson scatter, and using Figure \ref{fig:miscentering},
we expect the total scatter in richness to be 
$\sigma_{\ln \lambda|M} = 0.48 (0.23)$ for $\lambda = 30 (60)$.  
The corresponding scatter in mass
at fixed richness is obtained by multiplying the the slope of the mass--richness relation, which we estimate
in paper III as $\alpha=1.07$. Our final estimate for the scatter in mass at fixed
richness for $\lambda=30\ (60)$ is therefore $\sigma_{\ln M|\lambda}\approx 0.5\ (0.25)$.  
The scatter at $\lambda=60$ matches very well with out estimated scatter in mass from Figure B10 in paper III,
while the value obtained here for $\lambda=30$ is somewhat higher than that in Figure B10.  
Importantly, for $\lambda \gtrsim 60$, the centering probability $p$ increases relative to $\lambda=60$,
while $\sigma_R/R_c$ decreases.  Putting everything together, this suggests that miscentering of maxBCG
clusters is important for clusters with $\lambda \gtrsim 60$, at which point miscentering ``turns on'', and
leads to an increased scatter as a function of richness as one moves down in $\lambda$.  This feature
is indeed observed in Figure B10.  Note, however, that our estimate for the scatter in mass at 
$\lambda=30$ is somewhat higher than that of Figure B10, which suggests our miscentering model
has too many miscentered clusters at $\lambda=30$, and/or the miscentering kernel is too large for these
systems. 


\section{The Impact of Projection Effects on the Richness--Mass Relation}
\label{sec:proj}

\subsection{Projection Effects in High Density Regions}
\label{sec:correlated}

In section \ref{sec:results}, the galaxy density of non-cluster galaxies was modeled as a uniform density field where the mean density was set to the mean galaxy density 
over the entire sky.  In practice, however, clusters reside in high-density regions, so the mean local galaxy density of non-cluster galaxies 
$\bloc(i,g-r)$ is 
enhanced relative to global average.

\begin{figure}[t]
\epsscale{1.2}
\plotone{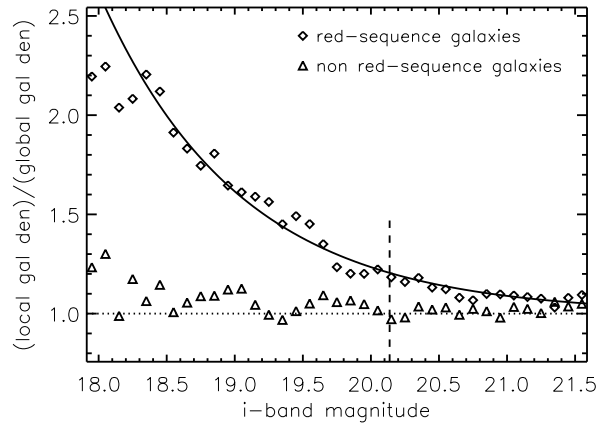}
\caption{Ratio of the mean galaxy density in annuli around maxBCG clusters in the
redshift range $z=[0.24,0.26]$ to the mean galaxy density of the universe.
We have split the galaxy sample into red-sequence and non red-sequence
galaxies since the two populations show very different enhancements.  The
fact that the ratio is always larger than one illustrates that clusters reside in
high density regions.}
\label{fig:local_boost}
\end{figure}

Figure \ref{fig:local_boost} shows the ratio of the local galaxy density $\bloc(i,g-r)$ to the global
mean $\bar b(i,g-r)$
as a function of magnitude for red-sequence (diamonds) and non-red sequence (triangles) galaxies.
The solid line is a fit to the red-sequence boost over the region $0.1L_* \leq L \leq L_*$.
The local galaxy density is estimated by
selecting the 2000 richest maxBCG clusters \citep[as determined using the richness estimator of][]{rozoetal09b}, 
and then stacking them in narrow redshift bins of width
$z=z_0 \pm 0.01$. 
Within each stack, we compute the mean galaxy density in an annulus of inner radius $R_{in}=1\ \hMpc$ and outer
radius $R_{max} = 2\ \hMpc$.  It is this galaxy density that we report as the local galaxy density $\bloc(i,g-r)$.  
Our choice of annulus is meant to purposely overestimate the mean galaxy density of non-cluster galaxies,
so that we may place a robust upper limit on the impact that this galaxy density boost can have on the richness--mass
relation.


We use the best-fit model from Figure \ref{fig:local_boost} as a new background model for our Monte Carlo simulations.
The exact model is
\be
\bmodel(i,g-r) = (1+ \avg{B|i} ) \bar b(i,g-r)
\ee
where
\be
\avg{B|i} = \avg{B_{20}}\exp(-0.95(i-20)) 
\label{eq:avg}
\ee
and $\avg{B_{20}}=0.28$.  Note that this model is fit over the luminosity range $[0.1L_*,L_*]$, but grows exponentially
fast with $i-20$.   In practice, the maximum boost we observe is $\approx 3$, so we impose a ceiling at $\avg{B|i}=3$
at very bright magnitudes.  The value of the ceiling has little impact on our results.

We find that the impact of this galaxy density boost to the richness--mass relation is unimportant.  The richness of galaxy clusters
is boosted by an average of 1.5 galaxies, corresponding to a $3\%$ bias for $\lambda=50$ clusters.  The scatter remains very
nearly Poisson, with $\Delta \sigma_{\lout}/\avg{\lout} \approx 1.3\%$.  These results are very sensible: the number of red-sequence
galaxies within our chosen aperture around a random piece of sky is one to a few.  Boosting this mean expectation by $30\%$
can add one or two more non-cluster galaxies to the richness estimate, but not much more than that. 
Thus, the fact that the mean galaxy density near galaxy clusters is higher than the global average
does not impact the richness--mass relation at a significant level.


\begin{figure}[t]
\epsscale{1.2}
\plotone{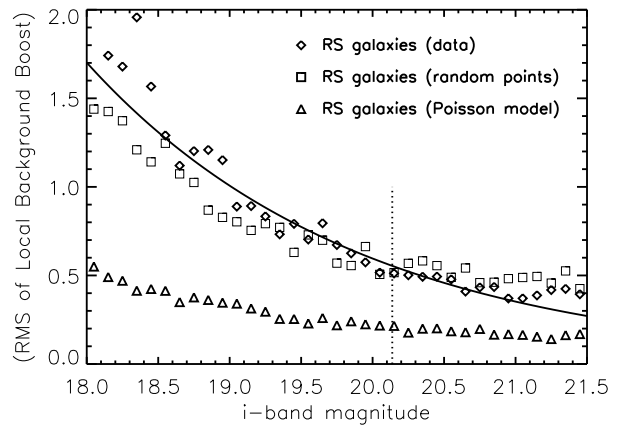}
\caption{RMS fluctuations in the local background of red sequence galaxies (diamonds) 
as a function of $i$-band magnitude for 
the redshift bin $z=[0.24,0.26]$.  The expected rms fluctuations from
a Poisson model in which every cluster explores the same boosted background density field 
is shown with triangles, while the fluctuations of the local density about random points are shown as squares.
The rms fluctuation in the local background is significantly larger than Poisson,
signaling that cluster-to-cluster variations in the density of non-cluster galaxies is highly significant.
Note, however, that these local fluctuations are only slightly larger than those about random points,
implying that correlated structures around galaxy clusters 
do not dominate the variance in the local cluster background.
The vertical dotted line corresponds to $0.2L_*$ at redshift $z=0.25$.
}
\label{fig:bgd_variance}
\end{figure} 


Note, however, that a clustered background implies not only that the mean galaxy density around clusters is larger than the global
average, but also that there can be cluster-to-cluster fluctuations in the background galaxy density.  
Figure \ref{fig:bgd_variance} shows the rms fluctuations of the local background boost
$B$ of red-sequence galaxies (diamonds) along with a best-fit model (solid line).
The rms fluctuations for blue galaxies are always smaller (not shown for clarity).  In all cases, we estimate
the rms fluctuation of the galaxy density field by estimating $B$ as in the previous section
using $100$ bootstrap resamplings of the
cluster catalog.  Also shown for reference is the rms fluctuation of red-sequence galaxies
about random points (squares), computed in the same way.  Because the number of random points
is significantly larger than the number of clusters available to us, one may worry that the additional
variance we observe is simply due to measurement error rather than clustering.  To test this hypothesis, 
we have measured the background fluctuations in Monte Carlo realizations from a model with a uniform
boosted galaxy density, which are significantly smaller than those for the observed cluster population.
Thus, we confirm that the galaxy density around galaxy clusters exhibits large cluster-to-cluster fluctuations.

 We incorporate these cluster-to-cluster fluctuations in our background model by setting
 \be
 \bmodel(i,g-r) = (1 + B(i) ) \bar b(i,g-r)
 \ee
 where $B$ is now a random function.  As in the previous section, we have ignored the color dependence of $B$, and we treat
 non red-sequence galaxies as red-sequence galaxies, which can only increase the impact of the variance of the density field.
 For simplicity, we also assume that the random fluctuations in $B$ at different magnitudes are
 perfectly correlated, so that large overdensities of galaxies in one magnitude imply a large overdensity of galaxies at all 
 magnitudes.  Not only do we expect this to be physically reasonable, it should also maximize the relative importance of projection effects
 as opposite fluctuations at different magnitudes cannot cancel each other out.
Our model for the function $B(i)$ is therefore 
\be
B(i) = \avg{B|i}+ \frac{\Delta B(i)}{\Delta B_{20}} \left( B_{20}-\avg{B_{20}} \right)
\label{eq:Bi}
\ee
where $\avg{B|i}$ and $\Delta B(i)$ are fits to the mean and standard deviation of the boost $B(i)$ of red-sequence
galaxies observed in the data, and $B_{20}$ is a random variable that determines the density boost of
$i=20$ red-sequence galaxies.   The function $\avg{B|i}$ is again given by equation \ref{eq:avg}, while for the standard deviation
we adopt 
\be
\Delta B(i) = \Delta B_{20} \exp\left[ -0.52(i-20)\right],
\ee
with $\Delta B_{20}=0.60$ as in the data.  The variable $B_{20}$ in equation \ref{eq:Bi} is modeled as a log-normal random
variable of the appropriate mean and standard deviation.


\begin{figure}[t]
\epsscale{1.2}
\plotone{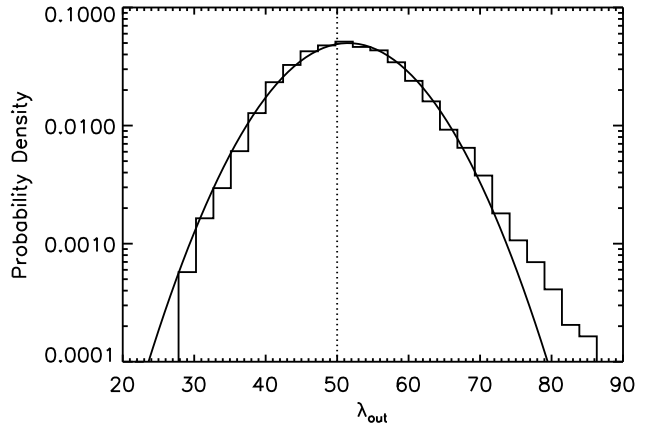}
\caption{Richness distribution of a cluster of input richness $\lin=50$ in the presence
of local background fluctuations.  The level of the local background fluctuations (both its mean 
and variance) has been
set to the values measured in the SDSS data (see text for details).  
The dashed curve is a Gaussian fit to the region $\lout\leq \lin+2\lin^{1/3}$, which has a normalization
constant $c=0.99$, implying $99\%$ of the cluster realizations have a scatter that is consistent with Poisson.
The non-gaussian tails reflect rare occurrences where the background is boosted to very high levels due
to random projection effects.
}
\label{fig:lambda_dist_varbgd}
\end{figure} 


The solid histogram in Figure \ref{fig:lambda_dist_varbgd} shows the $\lout$ distribution of $10^4$ realizations
of a cluster with $\lin=50$ with our lognormal local background model.  The distribution is clearly peaked
near the input value $\lin$, but is slightly biased, and exhibits a tail that extends to large $\lambda$ values,
which arise whenever the random background has an unusually large fluctuation due to random projection effects.
Also shown as a solid curve
is the best-fit Gaussian, where the fit is only performed over the region $\lout \leq \lin+2\lin^{1/2}$ in order to prevent
the tails of the distribution from affecting the fit.
The best-fit parameters of the Gaussian are
$\avg{\lout}=51.53\pm 0.08$ and $\sigma_{\lout}=7.92\pm 0.06$.  Once again, we find a $\approx 3\%$ bias on the
mean, while the width of the best-fit Gaussian remains close to the Poisson expectation ($\Delta\sigma_{\ln \lout} = 1.4\%$).
Importantly, however, the integral of the best-fit Gaussian is {\it not} unity.  Defining $c$ as the integral of the
best-fit Gaussian, we find $c=0.99$, implying that the tail of the distribution is formed by $1\%$ of the clusters.

The results described above for a $\lin=50$ clusters are generic.  We have repeated this experiment using clusters with $\lin=10,\ 20,$ and
$100$, and in all cases, $P(\lout|\lin)$ exhibited the same features: a main Gaussian peak whose mean is biased high, a width that is very close
to the Poisson expectation, and a normalization constant $c$ slightly less than unity.  As one might expect, the impact of a stochastic background
is smaller in richer systems.
Table \ref{tab:fits} summarizes those results.


\begin{deluxetable}{ccccc}
\tablewidth{0pt}
\tablecaption{Properties of Gaussian Fit to $P(\lout|\lin)$}
\footnote{Here, $\sigma_\lambda$ is the standard deviation of $\lambda$, as opposed to the 
standard deviation of $\ln \lambda$, which we use throughout most of our work.  The difference
is simply to allow us to be formally correct when we quote the relative the difference in terms of percentages
in the table.
The difference with respect to quoting $\Delta\sigma_{\ln \lambda}$ is insignificant.}
\tablehead{$\lin$ & 10 & 20 & 50 & 100 }
\startdata
$1-\avg{\lout}/\lin$ & $13.0\%$ & $6.8\%$ & $3.2\%$ & $1.9\%$ \\
$\avg{\lout}-\lin$ & 1.3 & 1.4 & 1.6 & 1.9 \\
$(\sigma_{\lout}-\sigma_{Poisson})/\avg{\lout}$ & $2.8\%$ & $2.2\%$ & $1.0\%$ & $0.5\%$ \\
c & $95.2\%$ & $97.6\%$ & $98.9\%$ & $99.3\%$
\enddata
\label{tab:fits}
\end{deluxetable}



\subsection{Cosmological Interpretation}
\label{sec:interpretation}

Our results paint a very clear picture of how projection effects operate in the real universe: for the vast majority of halos --- $\gtrsim 99\%$ at 
high mass and $\gtrsim 95\%$ at low mass (see table \ref{tab:fits}) ---
the density of non-cluster galaxies within the cluster field has a negligible impact on the richness estimate for the halo.  In roughly
$1\%-5\%$ of all cases, however, the halo resides within a galaxy density field that is much denser than average, which results in a large contribution
of non-halo galaxies to the richness estimate.  That is, $\lesssim 5\%$ of all halos suffer from projection effects.

\begin{figure}[t]
\scalebox{0.75}{\rotatebox{90}{\plotone{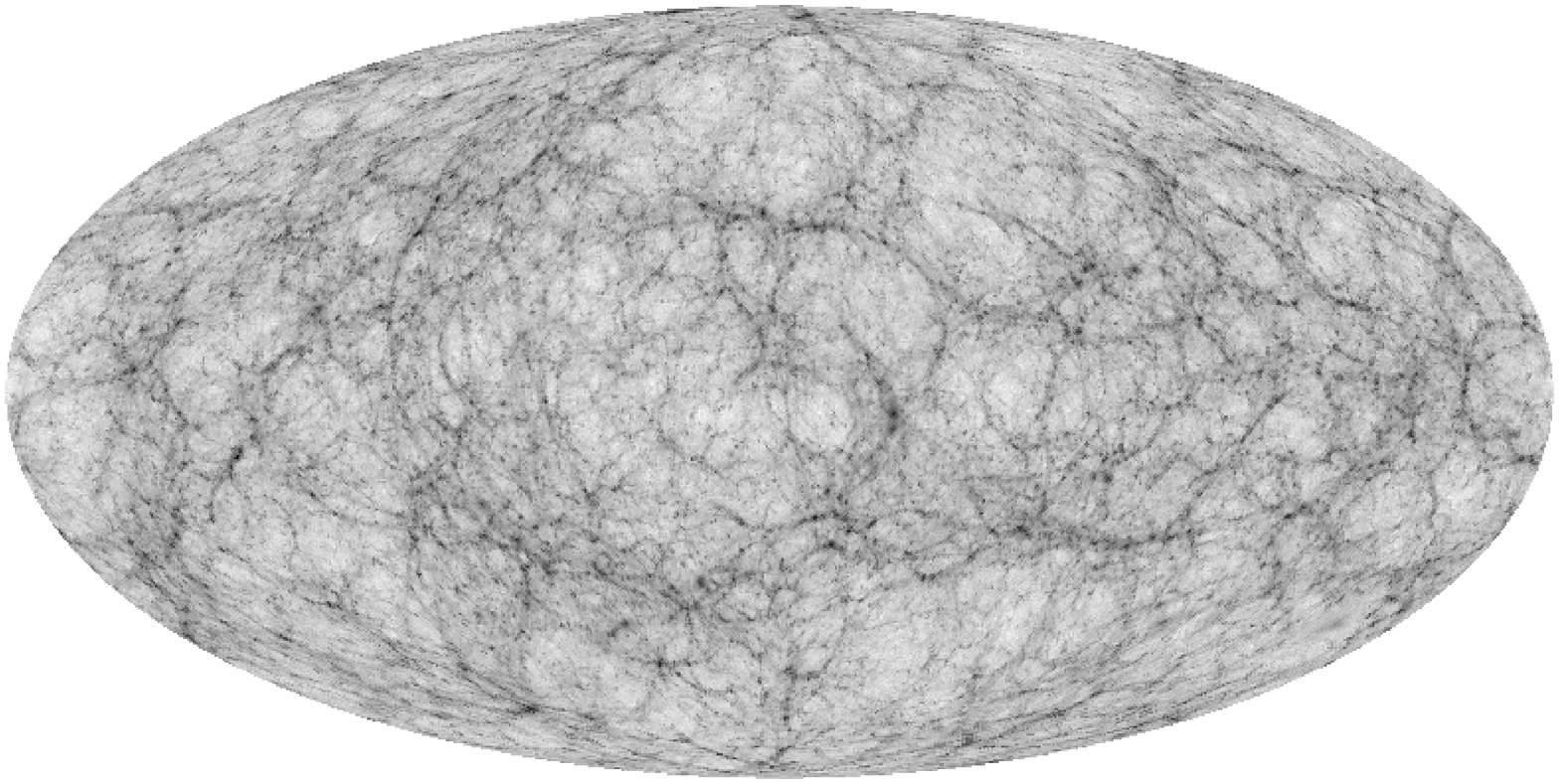}}}
\scalebox{0.75}{\rotatebox{90}{\plotone{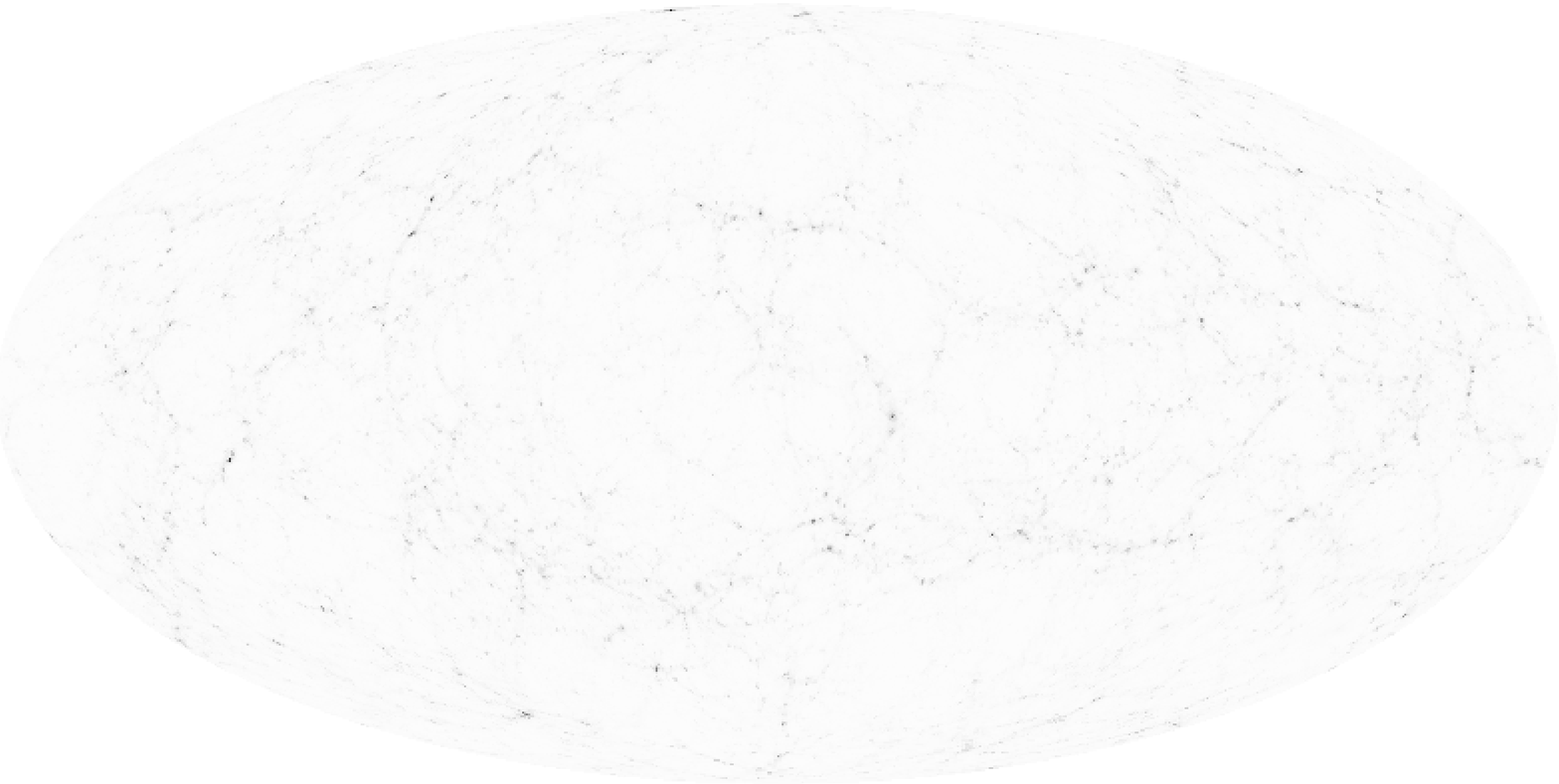}}}
\scalebox{0.75}{\rotatebox{90}{\plotone{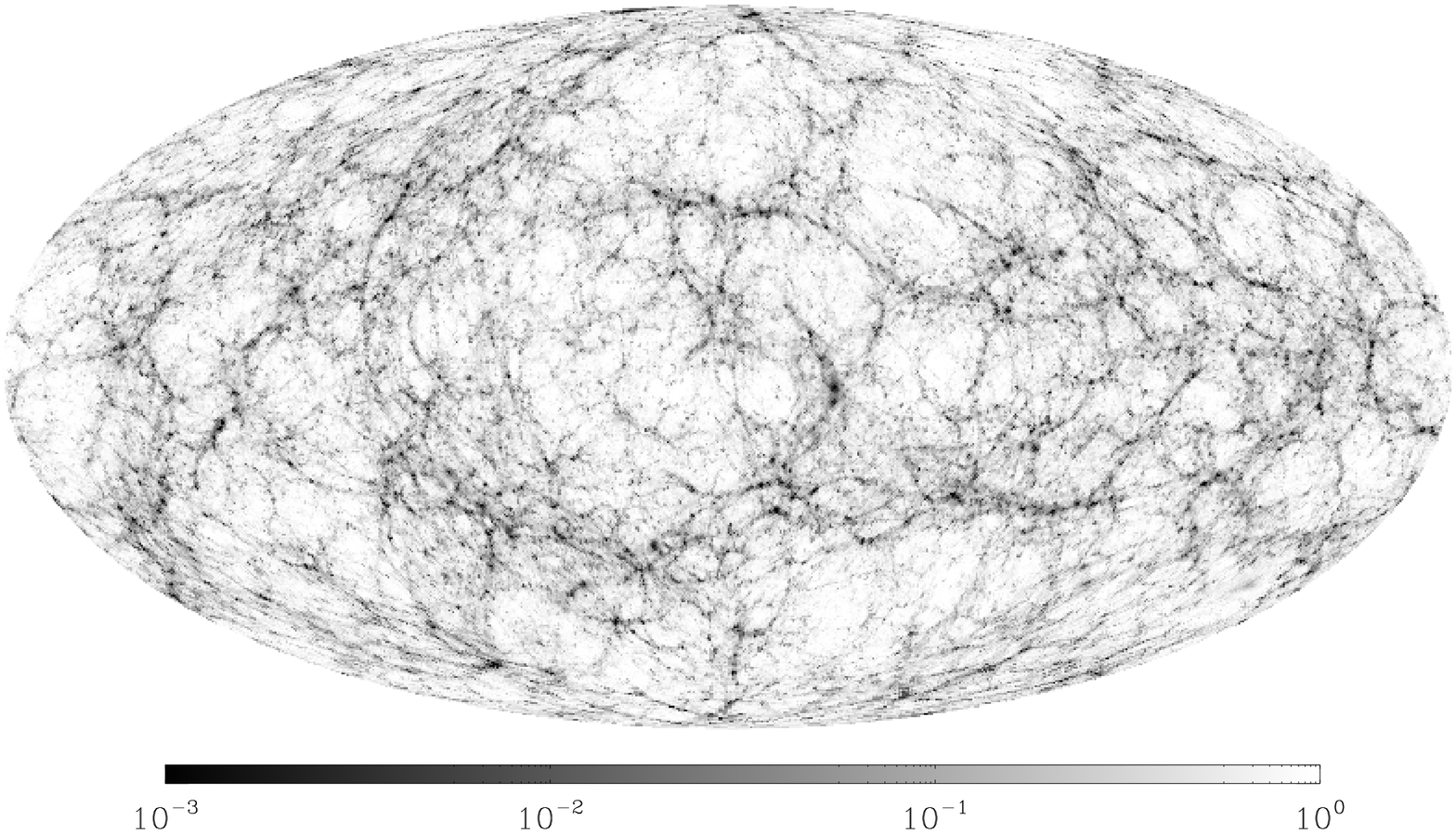}}}
\caption{The projected surface matter density around a random $z=0$, $M=1.57\times 10^{14}\ \msun$ halo in the Millenium 
simulation with gas \citep{springeletal05,Gazzola:2006p8392}.  
Only matter within a radius $R=170\ \Mpc$ of the halo is projected,  corresponding to the $2\sigma$ interval
spanned by the width of the red-sequence.
The density scale is such that darker regions correspond to higher density, and is logarithmic
in the top panel but linear in the middle panel.  Projection effects are linear in the density field, and therefore the
middle panel can be thought of as a map of how important local projection effects to that cluster are as a function
of line of sight.  Note most lines-of-sight fall in a white, empty space, for which projection effects are negligible.
To better illustrate this, in the bottom panel we should the same density field, but now color coded according
to the fractional area $f$, so that a pixel of density $\Sigma_{pix}$ is assigned a value $f$ that is the fraction
of the sphere covered by pixels with $\Sigma\geq \Sigma_{pix}$.  We see that all the dense knots from the top
and middle panels cover only $\approx 1\%$ of the sphere, while the filamentary structure covers $\approx 10\%$.
This illustrates that CDM cosmologies predict that most halos do not suffer from projection effects.
}
\label{fig:lss}
\end{figure}

Interestingly, this is exactly the kind of picture that one should expect from CDM cosmologies.
To see this, in the top and middle panels of
Figure \ref{fig:lss} we show the projected matter density in spheres of radius $R=170\ \Mpc$ around a 
random $10^{14}\ \msun$ (strictly speaking $M=1.57\times 10^{14}\ \msun$) halo in the Millenium Gas Simulation at z=0 
-- a GADGET-2-driven replica of the Millennium Simulation \citep{springeletal05}, 
where half of the one billion particles are treated as gas particles subject to hydrodynamics; for a further 
details see \citep{Gazzola:2006p8392, Hartley:2007p8444, Stanek:2010p27807}.  
The idea for these plots comes from the work of \citet{colbergetal99}, who used these
type of plots to to understand the large scale structure about galaxy clusters \citep{colbergetal99}.
The $170\ \Mpc$ radius is a very conservative estimate for
the distance along the line of sight corresponding to the width of the red-sequence, 
and the gray scale is chosen so that darker regions correspond to
higher densities, with the top panel being log-scale while the middle panel is linear.
These maps do not look qualitatively different when we use halos of higher and/or lower mass, nor when we vary 
$R_{max}$ in the range $100\ \Mpc-200\ \Mpc$.


Now, consider how an observer would see a cluster.  The line of sight from the observer to the cluster
pierces the density map
at exactly one point.  Because projection effects are linear in the density field,
the middle panel in Figure 9 can be thought of as a map of the importance of projection effects.  The fact
that this middle panel is essentially empty reflects the fact that 
for a uniform line of sight sampling, most observers will see a cluster with no projections.  Only a small subset of
observers will severely overestimate the richness, as we argued based on empirical evidence.

This is best illustrated by the lower panel in Figure \ref{fig:lss}, which was constructed as follows.  Starting
from the projected density field, for any given pixel with density $\Sigma_{pix}$, we compute the fractional area
of the sphere where $\Sigma \geq \Sigma_{pix}$.  Thus, strong overdensities have $f\lesssim 1$, while strong
underdensities have $f\approx 1$.  In this way, we can remap the density field from the top and middle panels
into a map of fractional area coverage.  We see that the tight density knots from the top and middle panels
cover only $\approx 1\%$ of the sphere, with filaments cover $\approx 10\%$ of the area, which is
consistent with our empirical estimates of the fraction of galaxy clusters that suffer from severe projection effects.

One question that remains to be addressed is whether or not projection effects are dominated by local structures
or uncorrelated structures.
To address this question, we compare the variance of the local density field around galaxy
clusters (Figure \ref{fig:bgd_variance}, diamonds) to the variance around random points in the sky (Figure \ref{fig:bgd_variance}, squares).
If correlated structures dominate the variance of the density field, then the former will be significantly larger than
the latter.  However, this is not what we observed, so we can conclude that while correlated structures certainly enhance
projection effects, they are not overwhelmingly dominant.

It is difficult to say whether this last conclusion --- that correlated structures do not necessarily dominate overall projection effects ---
generalizes to high redshift.  On the one hand, it is expected that the width of the red-sequence will correspond to increasingly
larger distances along the line of sight, which would tend to make uncorrelated structures more important.  On the other hand,
at fixed halo mass, higher redshift halos are rarer peaks, which would increase the relative importance of correlated structures.
Given the relatively modest impact that correlated structures have at low redshift, however, we would be surprised if projection
effects from uncorrelated structure ever becomes negligible.


\subsection{Projection Effects, and Their Implication for
Completeness and Purity}
\label{sec:comp_pur}

When interpreting our results, it is also important to distinguish the fraction of {\it halos} that suffer from projections effects, from
the fraction of {\it optically selected clusters} that suffer from projection effects.
Specifically, due to the steepness of the halo mass function, 
the frequency of optically selected clusters that suffer from projection effects will be higher, as at any richness there
are always more low mass halos scattering in and than out.  To estimate this effect, we proceed as follows.
First, we measure the richness 
$\lambda$ of every cluster in the maxBCG cluster catalog \citep{koesteretal07a}, and fit the corresponding abundance
function as a power-law, restricting ourselves to the richness range $30 \leq \lambda \leq 60$.\footnote{Here, $\lambda$ is 
measured as detailed in paper III, rather than using the simpler filters employed in this work.  Over the range of richness we 
consider, however, the two agree well.  We are not concerned about the detailed differences since our goal here is only to
provide a rough estimate of projection effects.}
We then select a fraction
$1-c(\lambda)$ of the clusters (as measured from  our simulations), boost their richness as described below, and 
recompute the richness function.  The difference between the original and boosted abundances gives the fraction
of clusters that suffer from projection effects.  We denote this fraction $p(\lambda)$, and refer to it as the purity.
For $c(\lambda)$, we fit the data in table \ref{tab:fits} to find
\be
c(\lambda) = 1 - 0.014(\lambda/40)^{-0.84}.
\ee

We consider three distinct methods for boosting the abundance of galaxy clusters, corresponding to an optimistic, a pessimistic scenario, and a super-pessimistic
scenario.
In the pessimistic scenario, we assume a halo of richness $\lambda$ is projected onto another halo of richness $\lambda$, so the boosted
richness becomes $2\lambda$.  This is pessimistic in that for unequal richness projections, the richer of the two objects can always be considered
the main halo.  The super-pessimistic scenario is as the pessimistic scenario, but we double the number of halos that suffer from projection
effects.
In the optimistic case, we simply demand that the projection effects be larger than Poisson fluctuations, so we set the
boosted richness to $\lambda+2\lambda^{1/2}$.  i.e. projection effects only increase the richness by twice the standard deviation
from Poisson statistics.  The motivation for relying on these simple models rather than our Monte Carlo simulations is
that our simulations are almost certainly not correct in detail, but these analytic arguments should bracket the correct answer.


\begin{figure}[t]
\epsscale{1.2}
\plotone{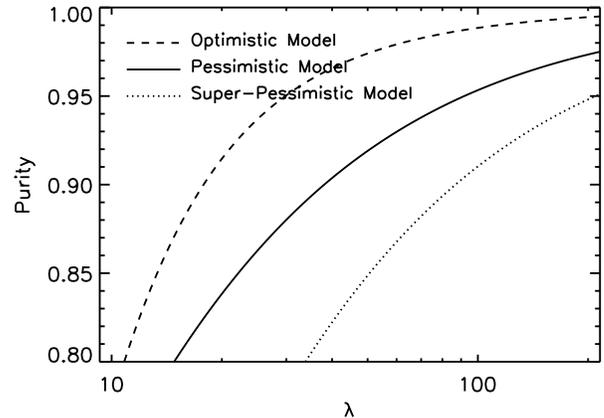}
\caption{The predicted purity of the optically selected cluster as a function of the richness $\lambda$.  By purity, we mean the fraction of clusters
of richness $\lambda$ that do {\it not} suffer from projection effects.  This quantity is computed using two simple models for projection effects,
an optimistic model (dashed) and a pessimistic one (solid).  The true impact of projection effects is likely to fall somewhere in between the two
lines, and probably closer to our optimistic model.  Thus, a reasonably value for the fraction of clusters that do not suffer from projection
effects for $\lambda\approx 30$ is $p=90\%$.
}
\label{fig:purity}
\end{figure} 


Figure \ref{fig:purity} shows the purity function for our three models.    At high richness $(\lambda \gtrsim 40)$, the purity
remains high $(p\gtrsim 0.9)$, but quickly decreases with decreasing richness.  
Note that by ignoring the curvature of the richness function, we are somewhat over-estimating the purity in the high richness
range, and under-estimating the purity in the low richness range.  In light of these considerations and the simple analytic nature
of our model, a reasonable value for the purity
function is $p\approx 90\% (95\%)$ at $\lambda=30\ (100)$.


\subsection{Discussion}
\label{sec:discussion}

Our results in this section are most directly comparable to the work of \citet{cohnetal07}
\citep[though see also][]{cohnwhite09,nohcohn11}
who relied on N-body simulations to explore the impact of local structures on richness estimates.
Their work is highly complementary to ours: N-body simulations provide a very realistic
treatment of correlated structures, whereas relying on Monte Carlo realizations of simple analytic
models allows us to generate many thousands of cluster realizations with varying sources of error,
a flexibility that was a necessary component of the study we undertook.
It should also be noted
 that the work of  \citet{cohnetal07} differs from ours
in that they employed a different richness estimator that is specifically tuned to their simulation, and, 
as noted earlier, such details will have a quantitative impact on our results.
Despite all these differences, the two results reach similar conclusions, with \citet{cohnetal07}
estimating that local projection effects are important for $\lesssim 10\%$ of all optically selected clusters.

It is not too surprising that this is the case.  To see this, we note that
that the density
boost due to cluster enviroment satisfies $\Delta B > \avg{B} \approx 0$.
The only way that this condition can be satisfied given the constraint $B>0$ is if the probability
distribution $\rho(B)$ is sharply peaked near $B\approx 0$, but has tails extending to large values of $B$.
As long as the tail is pronounced, differences in how one estimates the fraction of objects in the tail
will have little impact on the results.

Finally, we discuss the consequences of this work for the classic purity and completeness tests that are often applied to evaluate optical cluster 
finding algorithms.   The basic idea in these tests is this: to test completeness and purity, one takes the survey area, and randomizes the position and/or color
of every galaxy in the survey.   By embedding galaxy clusters at known positions and redshifts within the shuffled sky map, one can estimate
both purity and completeness.

There are multiple problems with the test we just described.
First, randomizing galaxy positions \citep[e.g.][]{postmanetal96,kepneretal99} is equivalent to setting the background galaxy density to a
uniform density field, which results in effectively no projection effects.
An alternative possibility is to randomize galaxy colors while
holding galaxy positions fixed or minimally displaced \citep[e.g.][]{gotoetal02,koesteretal07,bellagambaetal10}, but we have seen that
blue galaxies are significantly less structured than red galaxies, so once again this is not adequate.

The insight developed in this paper allows one to easily modify these tests to avoid such difficulties.  In particular,
given that projection effects by non-correlated structures appear to be at least comparable to projection effects from
correlated structures,
we expect that simply placing clusters in random points of the sky {\it without galaxy position and color randomization} should be a significantly
more realistic test of cluster completeness and the impact of projection effects \citep[this is similar to the approach adopted in][]{wenhanliu09}.
That said, such a test would still ignore the slight boost
to the mean and variance of the density field near clusters.  To fix this,
one could simply displace every cluster by two or three cluster radii, and re-estimate its richness.  Based on our results,
we expect this to be a significantly better test for evaluating the performance of cluster-finding algorithms.  Note that when performing such a test,
one will occasionally displace one cluster on top of another.  That is precisely the point: this type of overlap occurs in reality, so enforcing that
clusters not be displaced on top of one another, as is sometimes done, would lead to an under-estimate of projection effects.

Of course, one might wonder whether such tests are even necessary in the advent of large N-body simulations that can 
be populated with galaxies in ways that accurately reflect the known
universe \citep[see eg.][for works that have relied on numerical 
simulations to calibrate cluster selection]{milleretal05,rozoetal07b,dongetal08,milkeraitisetal10}.  
While such simulations are invaluable, and have, in fact, led to many improvements in our understanding
of optical cluster selection, the development of tests that can be applied on 
both on real and simulated data sets 
have the potential of being of critical importance.  Specifically, if various systematics can be calibrated directly from the
data using these empirically driven tests, then one can use simulations to test the efficacy of these methods rather than
having to rely on simulations for calibration purposes, which can introduce additional systematic uncertainties that are difficult to quantify.


\section{Summary}
\label{sec:summary}

We have used Monte Carlo simulations to explore various possible sources of {\it extrinsic} scatter of the richness--mass
relation in order to identify which, if any, are non-negligible.   By ``negligible'',
we mean that the source of noise under consideration does not increase the scatter of the richness--mass
relation by more than $5\%$ relative to the Poisson expectation.  This criterion is motivated by the precision required
to robustly estimate cosmological parameters in a DES-like survey (see Appendix \ref{app:scatter}).

We find that cluster-to-cluster variance in the properties of ridgeline galaxies, photometric errors,
and photometric redshift errors are all negligible.
Cluster triaxiality is border-line significant, increasing the scatter of the richness--mass relation by $\sim 5\%$ in the
richest systems.  The most significant source of extrinsic scatter by far is cluster miscentering, though
the details of the effect depends on the model parameters assumed.  Using the
 \citet{johnstonetal07} miscentering model, which is consistent with weak lensing observations \citep{ogurietal10}, 
 we find that miscentering is likely to be significant for $\lambda \lesssim 60$, but less so above this limit.
The total scatter from Poisson noise plus miscentering is close to the empirically estimated scatter
in Appendix B of paper III at $\lambda\approx 60$, but appears to over-estimate the scatter by $\lambda\approx 30$,
suggests that the miscentering model we considered over-estimates the importance of cluster miscentering
at low richness.  It is clear, however, that miscentering is likely to play an important role in the richness--mass
relation, particularly at low masses.

Finally, we considered the impact of projection effects on the richness--mass relation for a variety of assumptions
about the environment of galaxy clusters.  As one might expect, we find that clustering of the background density
field plays an important role on projection effects.  Less obvious however is that the boost to the mean galaxy density
around a cluster is unimportant.  Rather, projection effects are sourced by large cluster-to-cluster fluctuations in the
background galaxy density.   
Having estimated the variance of the background density field directly from SDSS data, 
we demonstrated that a small fraction of halos ($\approx 1\%-5\%$) are expected to
suffer from severe projection effects.
Due to the steepness of the mass function, the relative fraction of optically selected clusters 
that suffer from projection effects is higher, roughly i.e., $\sim 5\%-15\%$.  {\it We emphasize that all these
results are obtained using empirically motivated assumptions.}   
We also demonstrated that these results arise naturally in CDM cosmologies, and, indeed, our results 
are consistent with those \citet{cohnetal07}, who relied on N-body simulations to answer similar questions.

Our results have important consequences for the classical purity and completeness tests used to test optical cluster-finders.
Specifically, we demonstrated that simply randomizing the positions of galaxies in the sky and then inserting galaxy clusters
will grossly underestimate the importance of projection effects.
 A much better test is simply to place the clusters 
at random points in the sky without galaxy randomization, or, even better, to simply displace clusters by two to three cluster radii
and then to re-estimate their richness.  Most importantly, if such displacements lead to overlapping clusters, one should keep those
realizations, as such alignments naturally occur in the data set.

Overall, we believe the results from this work are very encouraging.  Between papers I and III, we believe we 
have considered all major modifications that could be made to our richness estimator. Consequently, we are confident
our final estimator is very close to optimal.  In addition, paper III also demonstrates that our estimator is extremely
robust.   With this work, we have developed a thorough qualitative understanding
of the sources of noise that can significantly impact the scatter of the richness--mass relation, and we have also been able to identify the most 
important observational systematic, namely miscentering.  Importantly, miscentering is not a problem of the richness estimator, but rather one having to
do with cluster finding: i.e. one needs to know how to adequately center clusters, a question that we have not addressed in this work.
Finally, we believe we have provided solid empirical evidence in favor of projection effects in galaxy clusters
being a relatively mild effect, and what's more, an effect that can be empirically calibrated.    We intend to return to the question of how
to model all these effects quantitatively in a future paper.   
All in all, we think these results justify being hopeful about our
ability to optimize cluster richness estimation for upcoming photometric cluster surveys, such as DES and LSST.

\acknowledgements ER would like to thank Joanne Cohn, Matthew Becker, and Andrey Kravtsov for helpful discussions 
and suggestions about the content of this manuscript.
ER is funded by NASA through the Einstein Fellowship Program, grant PF9-00068.  ESR thanks the TABASGO foundation.  
A.E.E. acknowledges support from NSF AST-0708150 and NASA NNX07AN58G. 
RHW and HW received supported from the DOE under contract DE-AC03-76SF00515.  
This work was supported in part by the Director, Office of Science, Office of High Energy and Nuclear Physics, of the 
U.S. Department of Energy under Contract No. AC02-05CH11231.

\newcommand\AAA[3]{{A\& A} {\bf #1}, #2 (#3)}
\newcommand\PhysRep[3]{{Physics Reports} {\bf #1}, #2 (#3)}
\newcommand\ApJ[3]{ {ApJ} {\bf #1}, #2 (#3) }
\newcommand\PhysRevD[3]{ {Phys. Rev. D} {\bf #1}, #2 (#3) }
\newcommand\PhysRevLet[3]{ {Physics Review Letters} {\bf #1}, #2 (#3) }
\newcommand\MNRAS[3]{{MNRAS} {\bf #1}, #2 (#3)}
\newcommand\PhysLet[3]{{Physics Letters} {\bf B#1}, #2 (#3)}
\newcommand\AJ[3]{ {AJ} {\bf #1}, #2 (#3) }
\newcommand\aph{astro-ph/}
\newcommand\AREVAA[3]{{Ann. Rev. A.\& A.} {\bf #1}, #2 (#3)}

\bibliographystyle{apj}
\bibliography{mybib}

\appendix

\section{How Precisely Must the Scatter be Known in a Cluster Counting Experiment?}
\label{app:scatter}

We wish to quantitatively define when a source of scatter is
observationally relevant.  To do so, let us assume that the total scatter in the richness--mass
relation of galaxy clusters is $\sigma_{tot}$, and that one performs a cosmological
analysis of cluster abundance data in which the model scatter $\sigma_{model}\neq \sigma_{tot}$.
Such an analysis will recover biased cosmological parameters, with the bias depending on
the difference between $\sigma_{tot}$ and $\sigma_{model}$.
Of course, if these biases are
small relative to the statistical uncertainties in the experiment,
then they are not observationally relevant.  Here, we consider deviations of the
total scatter $\sigma_{tot}$ from the model scatter $\sigma_{model}$ where the
model scatter is Poisson, and the total scatter includes extrinsic sources of
scatter such as those considered in the main body of this work.

We address this problem within the context of a DES-like cluster
cosmology experiment.  For our fiducial survey, we assume a survey
area of $5000\ \deg^2$, split into cells of $10\ \deg^2$ area each,
and a cluster selection threshold $M_{obs} \geq 7\times 10^{13}\
\msun$ over a redshift range $0\leq z \leq 1$.  We further assume the
cluster sample is binned in mass bins of width $\pm \Delta \log_{10}
M_{obs}=0.1$, corresponding to 5 bins per decade in mass, and we adopt
a log-normal model for $M_{obs}$--$M_{true}$ relation, with
\bea
\avg{\ln M_{obs}|M_{true}} & = & a + b \ln M_{true} + c\ln (1+z) \\
\sigma_{\ln M_{obs}|M_{true}} & = & \mbox{constant} = \sigma_{true}.
\eea
We assume both $M_{true}$ and $M_{obs}$ are measured in units of $7\times10^{13}\ \msun$, and set as our fiducial parameters
$a=c=0$ and $b=1$.   

Using the Fisher matrix technique described in detail in
\citet{wuetal08}, we then estimate what the cosmological constraints
derived from our fiducial cluster sample would look like, assuming
that the data is analyzed using the standard self-calibration
technique, and that the scatter is fixed a priori to a known value
$\sigma_{model}$.  We do not, however, enforce that $\sigma_{model}$ be
identical to $\sigma_{true}$.  Indeed, for any cosmological parameter
$p$, the two most important numbers that come out of the
\citet{wuetal08} Fisher matrix analysis are: a) the offset $\Delta p =
p_{obs}-p_{true}$ between the recovered value of the cosmological
parameter $p$ and its true value, and b) the estimated statistical
uncertainty $\sigma(p)$.  Thus, given any combination of
$\sigma_{true}$ and $\sigma_{model}$, we can estimate the ratio
$\Delta p/\sigma(p)$, and determine whether the difference between
$\sigma_{true}$ and $\sigma_{model}$ is observationally relevant or
not.


\begin{figure}[t!]
\plotone{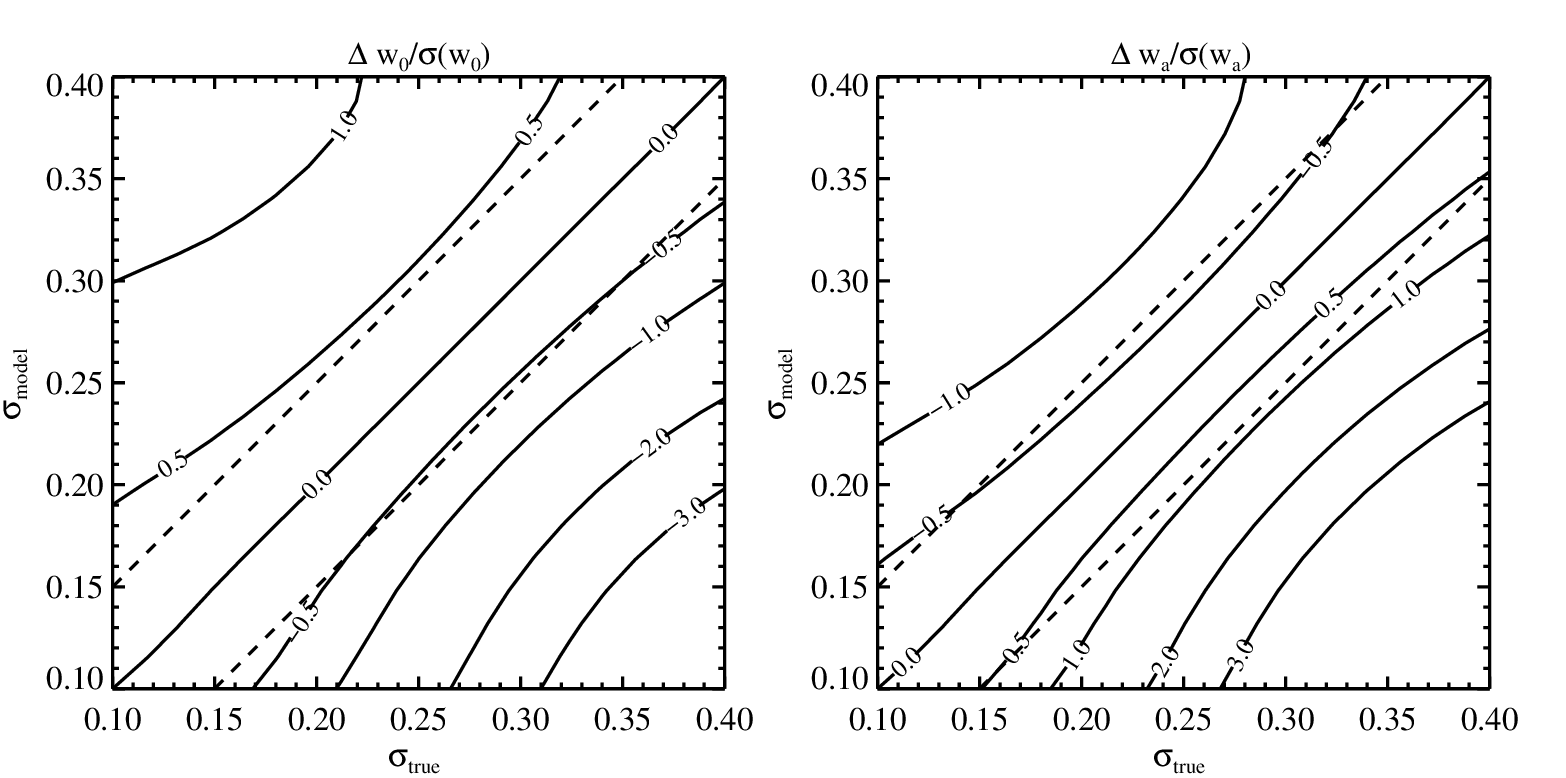}
\caption{Systematic errors in the inference of dark energy parameters
  $w_0$ and $w_a$ due to errors in modeling the scatter.  The
  contours and numbers show the systematic errors  $\Delta w_0$ ($\Delta w_a$) compared
  with statistical errors $\sigma(w_0)$ ($\sigma(w_a)$).  The y-axis $\sigma_{model}$
  indicates the scatter value we use in the analysis, while the x-axis
  $\sigma_{true}$ indicates the underlying true scatter.  As can be
  seen, if $\sigma_{true}-\sigma_{model} \geq 0.05$, the recovered dark
  energy parameters will be significantly biased.  
}
\label{fig:cosmology}
\end{figure} 


Figure \ref{fig:cosmology} shows contours of the ratio $\Delta
p/\sigma(p)$ for the dark energy parameters $w_0$ and $w_a$ as a
function of the scatter $\sigma_{true}$,
and the a priori scatter value $\sigma_{model}$ employed in the cosmological
analysis.  As is to be expected, the ratio $\Delta p/\sigma(p)$ goes to
zero as $\sigma_{model} \rightarrow \sigma_{true}$, and increases as the
difference between these two quantities increases.  The two diagonal
dash lines mark the equalities $\sigma_{model} = \sigma_{true} \pm 0.05$.  As
we can see, as long as $\sigma_{model}$ is within about $0.05$ of the
the true scatter $\sigma_{true}$, then the ratio $\Delta p/\sigma(p)
\lesssim 0.5$ for both $w_0$ and $w_a$.  In light of these
considerations, we define an extrinsic source of noise as
observationally relevant as one for which
$\sigma_{true}-\sigma_{model} \geq 0.05$.

\end{document}